\documentclass[%
reprint,
 prd,
superscriptaddress,
nofootinbib,
aps,
dvipsnames
]{revtex4-2}
\usepackage[utf8]{inputenc}
\usepackage{array,booktabs}
\usepackage{amsmath, amsthm, amsfonts,  amssymb, latexsym}
\usepackage{graphicx}
\usepackage{dcolumn}
\usepackage{bm}
\usepackage{hyperref}

\usepackage[mathlines]{lineno}
\usepackage{orcidlink}
\usepackage{breakurl}

\usepackage[normalem]{ulem}
\usepackage{xcolor}

\usepackage{enumerate}
\usepackage{mathrsfs}
\usepackage{graphicx}

\usepackage{subcaption}
\usepackage{xfrac}
\usepackage{lmodern}
\usepackage{multirow}

\usepackage{tikz}
\usetikzlibrary{angles,
                quotes}

\newcommand{\ud}{\mathrm{d}}
\newcommand{\Lie}{{\mathcal L}}

\newcommand{\emm}{m}
\renewcommand{\l}{\ell}

\newcommand{\Held}[1]{{#1}^{\circ}}
\def\tah{{\tau}^\circ}
\def\tabh{{\bar\tau}^\circ}
\def\tauH{\tau^{\circ}}
\def\taubH{\bar{\tau}^{\circ}}
\def\Psih{ {\Psi}^\circ}
\def\Psibh{ {\bar{\Psi}}^\circ}
\def\PsiH{\Psi^{\circ}}
\def\PsibH{\bar\Psi^{\circ}}

\def\rhopH{\rho^{\prime \circ}}
\def\rhopbh{{\bar\rho}^{\prime \circ} }

\def\Oh{{\Omega}^\circ}

\font\ec=ecrm0800 at 12pt
\def\thpH{\tilde{\th}'}
\def\edthH{\tilde\edth}
\def\edthpH{\tilde\edth'}
\def\th{\hbox{\ec\char'336}}
\def\edth{\hbox{\ec\char'360}}

\def\Dbar{\tilde \edth} % Held edth

\renewcommand{\S}{{\mathcal S}}
\newcommand{\E}{{\mathcal E}}

\renewcommand{\L}{{\mathcal L}}
\renewcommand{\O}{{\mathcal O}}

\newcommand{\sI}{{\mathscr I}}

\let\edth\relax

\font\ec=ecrm0800 at 12pt
\def\th{\hbox{\ec\char'336}}
\def\edth{\hbox{\ec\char'360}}

\def\Dbar{\tilde \edth}

\DeclareMathOperator{\thorn}{\text{\rm \th}}

\renewcommand{\Re}{\operatorname{Re}}
\renewcommand{\Im}{\operatorname{Im}}

\newcommand{\dd}{{\rm d}}

\def\non{\nonumber }
\def\half{\tfrac{1}{2}}
\newcommand{\Bco}[1]{{}_{#1}B_{\ell}}                 % Schwarzschild  {}_sB_l
\def\Dlw{D_{\ell\emm\omega}}                % denominator 2B 2Bbar + 9 M^2 w^2
\def\fHo{f^{\circ}}
\def\fbHo{\bar f^{\circ}}
\newcommand{\PhiN}[1]{\Phi^{#1\circ}}
\def\Phiadj{\Phi^{\rm adj}}

\def\zetal{\zeta^{\circ}_{l}}
\newcommand{\zetalM}[1]{\zeta^{\circ}_{l,#1}}
\def\zetan{\zeta^{\circ}_{n}}
\newcommand{\zetanM}[1]{\zeta^{\circ}_{n,#1}}
\def\zetam{\zeta^{\circ}_{m}}
\newcommand{\zetamM}[1]{\zeta^{\circ}_{m,#1}}
\def\zetamb{\bar\zeta^{\circ}_{m}}
\def\scri{\mathscr{I}^{+}}
\def\defeq{\mathrel{\mathop{:}}=}
\def\qb{\boldsymbol{q}}
\def\kb{\boldsymbol{k}}

\begin{document}
	
	\title{Metric reconstruction in a Bondi-like gauge}

    \author{Marc Casals}
\email{marc.casals@uni-leipzig.de}
\affiliation{%
	Institute of Theoretical Physics, Leipzig University, Br{\"u}derstra{\ss}e 16, 04103 Leipzig, Germany
}%
\affiliation{Scoil na Matamaitice agus na Staitistic\'i, An Col\'aiste Ollscoile, Baile \'Atha Cliath,  D04 V1W8, Ireland}
\affiliation{Centro Brasileiro de Pesquisas F\'isicas (CBPF), Rio de Janeiro, CEP 22290-180, Brazil}

		\author{Stefan Hollands \orcidlink{0000-0001-6627-2808}}
	\email[Contact author: ]{stefan.hollands@uni-leipzig.de}
	\affiliation{%
	Institute of Theoretical Physics, Leipzig University, Br{\"u}derstra{\ss}e 16, 04103 Leipzig, Germany
}%
\affiliation{%
	Max Planck Institute for Mathematics in Sciences (MiS), Inselstra{\ss}e 22, 04103
	Leipzig, Germany
}

\author{Vahid Toomani  \orcidlink{}}
	\email{vahid.toomani.2@uni-leipzig.de}
	\affiliation{%
    Institute of Theoretical Physics, Leipzig University, Br{\"u}derstra{\ss}e 16, 04103 Leipzig, Germany
	}

\author{Peter Zimmerman}
\email{zimmerator@protonmail.com}
	\affiliation{%
    Institute of Theoretical Physics, Leipzig University, Br{\"u}derstra{\ss}e 16, 04103 Leipzig, Germany
	}

	\begin{abstract}
    Metric perturbations obtained from a Hertz potential in an ingoing radiation gauge (IRG)
generically grow in an outgoing coordinate $r$ at future null infinity, and are therefore not asymptotically flat in the Bondi sense.  Hollands and Toomani
showed that this growth can be removed, within the
Green--Hollands--Zimmerman (GHZ) framework, by an IRG-compatible  ``residual'' gauge transformation together with a carefully chosen choice of the undetermined
integration ``constants'' of the Hertz potential. In this work we turn that proof into a
practical algorithm.  
We show that this entire completion is
generated by a single ``Held Master Scalar'' (HMS), 
which can be thought of as a function on future null infinity. The HMS is 
obtained from the leading asymptotic piece of the $0$-Weyl scalar through a Teukolsky--Starobinsky inversion. We give explicit formulas for the four integration constants of the Hertz potential associated with the $0$-Weyl scalar, and for the residual
gauge vector in terms of the HMS, thereby bringing the metric perturbation into a Bondi-like gauge. In Schwarzschild the entire construction is diagonal in 
$(\ell, m)$-modes and every completion field is available in
closed form; in Kerr the construction is tri-diagonal.  We have numerically implemented the full scheme in $(\ell,m)$ and in $m$-modes, including the corrector piece of the GHZ framework. We illustrate our scheme numerically for a point-particle on a circular orbit in a puncture scheme in Schwarzschild and in Kerr, by demonstrating the Bondi-like fall-off. We also find agreement with the known Detweiler redshift computations in Schwarzschild and Kerr. Our method directly gives the Bondi news, mass-, and angular momentum aspects, for which we give exemplary results in Kerr.
	\end{abstract}

	\maketitle

\section{Introduction}\label{sec:intro}

Large-mass-ratio inspirals of a comparatively small companion into a black hole present interesting conceptual challenges, such as the problem of scaling solutions of the Einstein equations, described from a physical perspective in \cite{Gralla:2008fg}, and more recently \cite{Hintz:2024dhy} in a limited scenario permitting complete mathematical control. Because such systems are among the principal targets of the Laser Interferometer Space Antenna \cite{AmaroSeoane:2012km} there is also a very strong practical motivation for developing accurate models capable of describing the dynamical evolution of the binary, and the emitted gravitational waves, over a very large number of orbits. The large mass ratios of up to $10^5-10^7$ expected in practice \cite{Babak:2017tow} are so far inaccessible to fully numerical schemes \cite{Lousto:2010ut}, in particular since a quite large number of such simulations corresponding to varying orbital- and black hole parameters would be needed for wave-form modeling \cite{Katz:2021yft,Chua:2020stf}. 

The current state of the art are instead perturbative, semi-analytical methods within the self-force scheme, see \cite{Barack:2018yvs,Pound:2021qin} as general references. In this scheme, the near field of the small object is typically isolated at various orders in perturbation theory---in practice, second order is considered necessary \cite{Hinderer:2008dm}. The residual field, i.e., the gravitation field with the small object field extracted, represents the large-scale ``trend'' of the gravitational field as opposed to the--often infinite--local perturbation, and is thought of as responsible for the self-force on the particle. This self-force gives rise to a gradual change of the orbital parameters, which must be updated throughout the dynamical evolution. Since the gradual change is very slow compared to the time-scale of the gravitational waves ($\gtrsim$ orbital period), it is possible to separate the problem into an evolution equation for the orbital parameters governed by a slow time, and perturbative calculations, at various orders, in a {\it fixed} Kerr background geometry and {\it fixed} orbital parameters, governed by a fast time \cite{Hinderer:2008dm,Barack:2018yvs,Wardell:2021fyy}.

In this paper, we focus on this second problem, which is the main input into the first. In each perturbative order, our problem, therefore, is basically to find a solution, $h_{ab}$, to the linearized Einstein equation ${\mathcal E}h_{ab} = T_{ab}$ on a fixed Kerr background, where $\E$ is the linearized Einstein operator\footnote{In our conventions, it is given by e.g., \cite[Eq. 26]{Hollands:2024iqp}.}, and $T_{ab}$ is the stress tensor at the respective order. In case $h_{ab}$ represents the residual field at that order, $T_{ab}$ will include, besides the point-particle stress tensor associated with the smaller object itself, also pieces arising from the lower order residual metric perturbations, and with the subtraction pieces of the near field (``punctures'' \cite{Pound:2013faa}), up to the same order. The latter are determined entirely locally 
\cite{Heffernan:2012vj,Heffernan:2012su,Bourg:2024cgh,Leather:2026gnc} 
and, while complicated, do not pose an essential problem. 

There are many different approaches to solving $\E h_{ab} = T_{ab}$ in practice. A common benchmark criterion is that one would like an $h_{ab}$ in a well-behaved gauge\footnote{The linearized Einstein operator is gauge invariant, so a gauge piece represents the part of the non-linear gauge transformation at the perturbation order that one is working in.}. Here, ``well behaved'' means e.g., that one would not like the metric to contain artificial singularities, or a non-decaying behavior at null-infinity, non-smooth behavior at the horizon, etc. In particular, it is highly preferable to find $h_{ab}$ in a Bondi-like gauge, such that asymptotic quantities like BMS-charges, memory, or fluxes are readily identifiable, or such that the contribution of $h_{ab}$ to the stress tensor at the next---in practice second---perturbation order is manageable.

In the present paper, we focus on the metric reconstruction method, originally due to \cite{Cohen:1974,Chrzanowski:1975wv} in the vacuum $(T_{ab}=0)$ case, but mention e.g., the recently developed Lorenz gauge method \cite{Dolan:2023enf,Dolan:2013roa} or \cite{Li:2026rkf} as interesting alternatives. Even though, as our application, we will consider a concrete puncture scheme yielding the residual part of the metric perturbation from an ``effective'' stress energy tensor $T_{ab}$ at first order, and for a circular equatorial orbit of Kerr, our scheme is general and conceptually independent of such restrictions as long as $T_{ab}$ has sufficient, and standard for radiating sources, decay \eqref{Tdec} at future null infinity. 

The broad aim of reconstruction methods is to obtain the metric perturbation $h_{ab}$ from the spin $s=+2$ Weyl scalar $\psi$, which obeys Teukolsky's decoupled separable equation for each mode~\cite{Teukolsky:1973ha}. Such mode decompositions are particularly suitable for systems with periodic sources representing orbiting point-like companions. From $\psi$, one then obtains a so-called Hertz potential, $\Phi$, by a well-known and easy inversion \cite{Ori:2002uv}, and then the metric perturbation as $h_{ab} = 2\Re \S^\dagger_{ab} \Phi$, where $\S^\dagger_{ab}$ is a tensor-valued ``reconstruction operator'' \cite{Cohen:1972pv,Chrzanowski:1975wv}. Actually, this scheme by itself is inconsistent for generic non-trivial sources $T_{ab}$. But as Green--Hollands--Zimmerman (GHZ) showed \cite{Green:2019nam}, it can be adapted by adding a correction piece, $x_{ab}$ determined by transport equations, to $2\Re \S^\dagger_{ab} \Phi$. A practical way to combine the GHZ scheme with the ``no-string-gauge'' methods by \cite{Pound:2013faa,Merlin:2016boc}, for determining the residual metric perturbation, was proposed later in \cite{Toomani:2021jlo}, and implemented for a circular orbit in Schwarzschild in \cite{Bourg:2024vre}. While the approach \cite{Toomani:2021jlo} avoids a stringlike gauge singularities emanating from the source, it does not by itself produce a metric perturbation in an asymptotically flat gauge. 

While far from obvious, it is physically plausible that $h_{ab}$ actually {\it has} an asymptotically flat gauge representative. However, even if it does (which we assume is the case and provide evidence for\footnote{For solutions of compact support (in some gauge), this follows from non-trivial mathematical theorems about the falloff of solutions to the linearized Einstein equation. However, these theorems do not directly apply to the retarded-type solutions under consideration here, wherein a source extending all the way to past time-like infinity is present.}) it is non-trivial to actually find this gauge representative, see \cite{Leather:2026zhl,Spiers:2026yqx} for recent works investigating and using asymptotically flat gauges in a self-force setting. In this paper, we instead follow a recent extension of the GHZ metric reconstruction method due to \cite{Hollands:2024iqp}, which delivers a gauge vector field $\zeta^a$
({\it and at the same time} a particular choice of the non-unique Hertz-potential $\Phi$) achieving a {\it Bondi-like gauge}. The total metric perturbation is hence
\begin{equation}
\label{eq:GHZdecomp}
    h_{ab} = 2\,\Re \S^\dagger_{ab} \Phi + x_{ab} -\L_\zeta g_{ab} + \dot g_{ab}
\end{equation}
wherein $g_{ab}$ is the Kerr metric, and $\dot g_{ab}$ a metric perturbation corresponding to an infinitesimal change\footnote{$(\delta M, \delta a)$ may be found from $T_{ab}$, as described in \cite{Toomani:2021jlo}. $\dot g_{ab}$ may be presented in IRG \cite[App. C]{Hollands:2024iqp}.} $(\delta M, \delta a)$ of the Kerr parameters $(M,a)$. 
The reconstructed piece $2 \Re \S^\dagger_{ab} \Phi$ and the corrector $x_{ab}$ by design are in ingoing radiation gauge (IRG)---meaning a vanishing contraction with the outgoing principal null direction $l^a$---while 
$\dot g_{ab}$ may be presented in IRG by a gauge choice \cite[App. C]{Hollands:2024iqp}. Because $\zeta^a$ is a ``residual'' i.e., IRG compatible gauge vector field \cite{Price:2006ke}, the {\it entire} expression \eqref{eq:GHZdecomp} therefore is in IRG. Our Bondi-like gauge means this, {\it plus} the remaining four decay conditions \eqref{eq:hdec} on the null-tetrad components of $h_{ab}$.

Building on \cite{Hollands:2024iqp}, we will present a remarkably simple algorithm for obtaining {\it both} $\zeta^a$, {\it and} the particular choice of $\Phi$ at the same time. 
This algorithm is as follows:
\begin{enumerate}[(i)]
\item 
Find the leading order 
asymptotic piece $\psi^{5\circ}$ of the ``0'' perturbed-Weyl-scalar component $\psi$ (the $r^{-5}$ piece), which is used to 
\item 
Find a certain ``{\it Held Master Scalar}'' (HMS) $f^\circ$, where a $\circ$ indicates a quantity that is intrinsic to future null infinity, or more pragmatically, a function of $u,\theta,\varphi$ in an outgoing Eddington-like coordinate system. $f^\circ$ is determined from $\psi^{5\circ}$ by an angular Teukolsky-Starobinsky relation. 
\item 
Apply certain explicit geometrical differential operators to our HMS $f^\circ$, to obtain both the leading pieces $\Phi^{3\circ}, \dots, \Phi^{0\circ}$ (the coefficients of the $r^3, \dots, r^0$-pieces of $\Phi$, which are undetermined from the inversion relation linking $\psi$
and $\Phi$), and the components of the residual gauge vector field $\zeta^a$.
\end{enumerate}
Since the corrector $x_{ab}$ can be found by three simple integrable \cite{Toomani:2021jlo} transport equations \cite{Green:2019nam} with explicit Green's function \cite{Casals:2024ynr} if we want, since it is known how to solve Teukolsky's equation for $\psi$ with retarded boundary conditions, and since it is known \cite{Ori:2002uv} how to solve the inversion problem $\psi \to \Phi$, the problem of finding $h_{ab}$ as a solution to $\E h_{ab} = T_{ab}$ in reconstructed form {\it and} Bondi-like form near future null infinity is solved by (i)---(iii). 

We can consider both a so-called ``$m$-mode'' scheme (factoring out only the 
$e^{-i\omega t+im\varphi}$-dependence of the solutions), or a full ``$(\ell,m)$-mode'' scheme (factoring out also the spheroidal harmonics) for (i)---(iii). 
In general, our implementation architecture supports both schemes, though we mostly describe the $m$-mode scheme in Kerr. As an exception from this, we use an $(\ell,m)$-mode scheme for solving the Teukolsky equation for $\psi$. But we note that also this step could be based entirely on an $m$-mode scheme once $\psi$ is given as a numerical solution on an $(r,\theta)$-grid covering the exterior of Kerr, e.g., using the very recently developed spectral method by \cite{Cownden:2026msq}. As a consistency check of our scheme, we numerically check that $h_{ab}$ is in a Bondi-like gauge, and we compute the Detweiler redshift quantity.

Expectedly, the $(\ell,m)$-mode scheme is particularly straightforward in Schwarzschild-spacetime, but we also outline the necessary changes in Kerr. These are mainly due to the fact that the angular GHP-like operators do not act as simple raising or lowering operators, unlike Schwarzschild \cite{Chandrasekhar:1983}. We have successfully performed consistency checks of some of our main numerical findings between the $m$-- and $(\ell,m)$ mode schemes.

The concrete proof-of-concept type application of our general method that we consider is to find the first order residual metric perturbation $h_{ab}^{\rm R}$ within a puncture scheme \cite{Pound:2013faa}, for circular equatorial orbits $\gamma$ in Kerr. Here, the concrete problem to be solved is $\E h^{\rm R}_{ab} = T^{\rm E}_{ab}$, where
$T_{ab}^{\rm E}$ is an effective source taking into account the original point particle source, 
\begin{equation}
\label{eq:Tabdef}
    T_{ab}=\mu\int \ud T \, u_a u_b \, \delta^4[ \, . \, - \gamma(T)]/\sqrt{-g},
\end{equation}
($u^a = \dot \gamma^a$ is the unit tangent)
and additionally a stress tensor corresponding to the singular field near the body, called a ``puncture'', $h^{\rm P}_{ab}$. In formulas,
\begin{equation}
\label{eq:TEdef}
    T_{ab}^{\rm E} := T_{ab} - \E h^{\rm P}_{ab}.
\end{equation}
In concretely constructing $T^{\rm E}_{ab}$ in our setting, we partly rely on previous non-trivial works   \cite{Bourg:2024cgh,Bourg:2024vre}.
The full metric perturbation at first order would then be $h_{ab}=h^{\rm P}_{ab} + h^{\rm R}_{ab}$, where $h^{\rm R}_{ab}$ would e.g., be used to compute the self-force near the particle. Although we could do this, we will instead mostly concentrate on the asymptotic region where the technical advance of our new method plays out. 

We finally note that, if one can solve the non-trivial problem of constructing $T^{\rm E}_{ab}$ at higher orders in perturbation theory, which includes the problem of punctures at higher orders \cite{Pound:2021qin}, our method and $m$-mode implementation pipeline would directly generalize to this case: We would again find $\psi$ using a some Teukolsky solver, then fit the inversion relation to get $\Phi$, then use our integration implementation to find $x_{ab}$, and then go through the same steps (i)---(iii), which does not use particular features about the orbit or perturbation order. 

\section{Conventions for Kerr background and differential operators}

 Our conventions regarding the Kerr-metric, the Newman-Penrose (NP) \cite{Chandrasekhar:1983} -- and Geroch-Penrose-Held (GHP) formalisms \cite{Geroch:1973am}, are the same as e.g., in ~\cite{Green:2019nam}, \cite{Hollands:2024iqp}, where one can also find reviews of these formalisms. In particular, the signature of the metric is $(+,-,-,-)$. As in these references, we will actually mainly use Held's modified GHP calculus~\cite{Held:1974,Held:1975}, in the same conventions as stated in~\cite{Hollands:2024iqp}. 
 The reader is not expected to have any familiarity with these formalisms since for our concrete application, it is only necessary to recall the definitions of certain coefficient functions and differential operators. 

These are stated in the Kinnersley frame; see, e.g., ~\cite{Green:2019nam}, \cite{Hollands:2024iqp}, and Boyer-Lindquist coordinates called $(t,r,\theta,\varphi)$, or in outgoing Eddington coordinates called $(u=t-r_*,r,\theta,\varphi_*)$ for Kerr, which are related by
\begin{equation}
\ud r_* = \frac{r^2+a^2}{\Delta} \ud r, \quad
\ud\varphi_* = \ud\varphi - \frac{a}{\Delta} \ud r, \quad
\end{equation}
where $\Delta=r^2-2Mr+a^2.$
In these conventions, the coefficients that we require, are
\begin{equation}\label{eq:rho def}
\rho = -\frac{1}{r-ia\cos\theta},
\end{equation}
$\PsiH, \rhopH, \tauH, \Oh$, and a certain geometric basis of directional derivative operators, see Table \ref{tab:Held-coeffs-Kerr}. 

{\it Held's $\circ$-notation:} As a matter of notation, in the Held-formalism, a $\circ$ on a quantity $X^\circ$ intrinsically means a GHP scalar that is annihilated by the GHP-operator $\th$; pragmatically this just means that it is independent of $r$ in Eddington coordinates in our conventions. By construction~\cite{Held:1974,Held:1975}, any Held operator- or NP coefficient in Table \ref{tab:Held-coeffs-Kerr} preserves this property, {\it unlike} the usual GHP operators.

\begin{table*}[t]
\caption{Summary of Held's GHP quantities in outgoing Eddington
  coordinates $(u,r,\theta,\varphi_*)$ and the Kinnersley frame
  (taken from~\cite{Green:2019nam}).
  On a mode with dependence $e^{-i\omega u + im\varphi_*}$
  (or with $e^{-i\omega t + im\varphi}$ in Boyer--Lindquist coordinates),
  one puts $\partial_u \to -i\omega$ and $\partial_{\varphi_*} \to im$.
  $\edthH$ and $\edthpH$ are thereby equal to the spin-raising and
  spin-lowering operators when acting on spin-weighted spherical harmonics;
  see, e.g.,~\cite[App. C]{Toomani:2021jlo}.}
\label{tab:Held-coeffs-Kerr}
\begin{ruledtabular}
\renewcommand{\arraystretch}{1.6}
\begin{tabular}{c >{$\displaystyle}l<{$} c}
  \text{Held quantity}
    & \multicolumn{1}{c}{\text{Value}}
    & \text{Spin weight} \\
  \hline
  $\tau^\circ$
    & -\frac{ia\sin\theta}{\sqrt{2}}
    & $+1$ \\
  $\rho^{\prime\circ}$
    & -\frac{1}{2}
    & $\phantom{+}0$ \\
  $\Psi^\circ$
    & M\rule{0pt}{2.8ex}
    & $\phantom{+}0$ \\
  $\Omega^\circ$
    & -2ia\cos\theta \rule{0pt}{2.8ex}
    & $\phantom{+}0$ \\
  $\thorn$
    & \partial_r\rule{0pt}{2.8ex}
    & $\phantom{+}0$ \\
  $\thpH$
    & \partial_u
      - \frac{1}{2}\,
        \frac{r^2 - 2Mr + a^2}{r^2 + a^2\cos^2\!\theta}
      \,\partial_r
    & $\phantom{+}0$ \\
  $\edthH$
    & -\frac{1}{\sqrt{2}}\!\left(
        \partial_\theta
        + i\csc\theta\,\partial_{\varphi_*}
        + ia\sin\theta\,\partial_u
        - s\cot\theta
      \right)
    & $+1$ \\
  $\edthpH$
    & -\frac{1}{\sqrt{2}}\!\left(
        \partial_\theta
        - i\csc\theta\,\partial_{\varphi_*}
        - ia\sin\theta\,\partial_u
        + s\cot\theta
      \right)
    & $-1$
\end{tabular}
\end{ruledtabular}
\end{table*}

The quantity $s$ is the spin weight of a GHP scalar or operator \cite{Geroch:1973am}. It must be taken into account for every quantity these operators are acting on, and a GHP operator with the indicated weight will shift these by the amount associated with the operator. A nice feature of the above Held operators-- and scalars is that they lead to manifestly gauge invariant equations, which thereby could be converted readily from the Kinnersley frame to any other one by a simple local boost/rotation factor associated with the NP tetrad change, see e.g., \cite{Hollands:2024iqp}. This would be practically useful if we wanted to convert our abstract formulas e.g., to the Hartle-Hawking-- or Carter frames, and is also of course geometrically much more satisfying.

\section{The asymptotically flat reconstruction scheme}
\label{sec:scheme}

Here we describe our asymptotically flat reconstruction scheme for obtaining \eqref{eq:GHZdecomp}. It is extracted from \cite[Thm. 4]{Hollands:2024iqp} and its derivation in \cite[App. J]{Hollands:2024iqp}. The theorem applies rigorously to (smooth) solutions to the linearized Einstein equations (EE) $\E h_{ab} = T_{ab}$, under the following 

{\it Assumptions:}
\begin{enumerate}
    \item The NP-components of the stress energy tensor $T_{ab}$ in our Kinnersley frame has the standard\footnote{
    Regarding our condition for $T_{ll}$, we are one order of $1/r$
    more generous than \cite{Hollands:2024iqp}. It can be seen that all the relevant proofs of \cite{Hollands:2024iqp} still go through under this condition. 
    } behavior
\begin{equation}\label{Tdec}
\begin{split}
T_{nn} =&  \ O\left(\frac{1}{r^2}\right) , \\
T_{ln}, T_{m\bar m}, T_{nm} =& \ O\left(\frac{1}{r^3}\right), \\
T_{lm}, T_{mm}, T_{ll} =& \ O\left(\frac{1}{r^4}\right), 
\end{split}
\end{equation}
towards ${\mathscr I}^+$, i.e.,
as $r \to \infty$, at fixed $(u,\varphi_*,\theta)$.
The conditions are equivalent to the 
Bondi coordinate-type conditions given e.g., in \cite{Flanagan:2015pxa}.

\item There exists {\it some} gauge such that
the NP components of $h_{ab}$ have asymptotic expansions 
with decay orders 
\begin{equation}
\label{eq:hdec}
\begin{split}
    h_{ll}, h_{lm}, h_{ln}, h_{m\bar{m}} &= O\left(\frac{1}{r^2}\right), \\
h_{nn}, h_{nm}, h_{mm} &= O\left(\frac{1}{r}\right),
\end{split}
\end{equation}
as $r \to \infty$, at fixed $(u,\varphi_*,\theta)$.
These conditions would be satisfied if one were to assume that there exists a Bondi-gauge in the usual (coordinate) sense for $h_{ab}$, see \cite[App. D]{Hollands:2024iqp}. In an IRG, all NP components involving at least one ``$l$'' would vanish.
\item There exist $\delta M, \delta a$ such that, subtracting a corresponding zero mode in IRG satisfying \eqref{eq:hdec} \cite{Toomani:2021jlo} (see, e.g., \cite[Eq. C.1]{Hollands:2024iqp} for a concrete formula)
\begin{equation}
\label{zerosubtract}
    h_{ab} \to h_{ab} - \dot g_{ab},
\end{equation}
the leading expansion coefficients in the asymptotic expansion all go to zero as $u \to -\infty$.
\item At ${\mathscr I}^+$ the Bondi news tensor satisfies the usual Price decay law \cite{Price1972b} as $u \to \infty$.
\end{enumerate}

{\it Theorem:} \cite[Thm. 4]{Hollands:2024iqp} There exists a gauge in which $h_{ab}$ takes the form 
\eqref{eq:GHZdecomp}, and such that it is in addition in a Bondi-like gauge.

{\it Bondi-like gauge:} This means that $h_{ab}$  is in IRG, i.e. 
$h_{lm}=h_{ln}=h_{ll}=0$, {\it and} that the remaining four asymptotic behaviors in \eqref{eq:hdec} hold for the non-zero NP components. This gauge could easily be converted to an actual Bondi gauge in the traditional (coordinate) sense \cite[App. D]{Hollands:2024iqp}, but for the purposes of extracting quantities associated with gravitational radiation, or for using $h_{ab}$ to build the source at the next perturbative order, the Bondi-like gauge appears just as good if not better.

Of course in practice, one does not know $h_{ab}$ from the outset, so one does not know whether conditions 2)--4) on the metric perturbation hold unless one can appeal to abstract mathematical theorems to this effect! We are unfortunately not aware of such abstract theorems in the specific self-force settings below, mainly due to the fact that the particle exists forever. We therefore need to {\it assume} that the conclusions hold. Then, we can extract the following {\it prescription} from \cite[Thm. 4]{Hollands:2024iqp}, to actually {\it find} the pieces $x_{ab}, \Phi, \zeta^a, \dot g_{ab}$ in the decomposition \eqref{eq:GHZdecomp} for $h_{ab}$.

{\it Algorithm:}
Step 1) The perturbed Teukolsky scalar\footnote{This $\psi$ is $-\psi_0$ of \cite{Hollands:2024iqp} and many other references; we have no need for $\psi_4$ in this article, and consequently we drop the subscript ``0''.} 
\begin{equation}
\label{eq:psidef}
\psi \equiv  \delta C_{lmlm}
\end{equation}
of spin-weight $s=+2$, is determined through Teukolsky's equation \cite{Teukolsky:1973ha}, 
\begin{equation}
\label{eq:Teuksource}
    \O \psi = \S^{ab} T_{ab}, 
\end{equation}
where $\S^{ab}$ is Teukolsky's source operator, and $\O$ is Teukolsky's wave operator, see App. \ref{app:Teuk} for their concrete expressions in terms of our operators in Table \ref{tab:Held-coeffs-Kerr}. We assume {\it retarded} boundary conditions\footnote{This is not critical for the scheme; in fact \cite{Hollands:2024iqp} assumed initial conditions on some hyperboloidal slice.}. Then by the peeling theorem\footnote{This would follow in a rigorous setting by \eqref{eq:hdec}; here it is strictly speaking an assumption which we will find to be correct in our application to self-force.} 
\begin{equation}
\label{eq:peeling}
    \psi = \frac{\psi^{5\circ}}{r^5} + O\left(\frac{1}{r^6}\right).
\end{equation}

Step 2) The Hertz potential $\Phi$ is obtained from the well-known \cite{TeukolskyPress1974} inversion relation\footnote{This equation refers to Eddington coordinates $(u,r,\varphi_*,\theta)$ and the Kinnersley frame. The GHP invariant form of this equation is $\th^4 \bar\Phi = 2\psi$. Note that this $\Phi$ is $-\Phi^{\rm ht}$
e.g., in \cite{Hollands:2024iqp}.}
\begin{equation}
\label{eq:invrel}
\partial_r^4 \bar\Phi = 2\psi.  
\end{equation}
We then have a unique {\it minimal solution}, i.e., such that $\Phi^{\rm min} =O(r^{-1})$. 
However, this minimal solution is {\it not} the one appearing in 
the asymptotically flat IRG metric \eqref{eq:GHZdecomp}\footnote{Also, neither $\Phi^{\rm min}$ nor $\Phi$ as in \eqref{eq:Phi completion intro} is equal to the Hertz potential $\Phi^{\rm ns}$ appearing in the no-string construction of \cite{Toomani:2021jlo}. That object would satisfy, e.g., $\O^\dagger \Phi^{\rm ns} =0$ in the source free region, but not ours.}. Instead, we have to add to the minimal solution 
specific terms called {\it completion pieces} $\Phi^{\rm adj}$:
\begin{subequations}
\label{eq:Phi completion intro}
\begin{align}
   \Phi^{\rm min}
\longrightarrow
\Phi &\equiv \Phi^{\rm min} 
+ \Phi^{\rm adj}\\
\Phi^{\rm adj} &\equiv \PhiN{0}
+
\frac{\PhiN{1}}{\rho}
+
\frac{\PhiN{2}}{\rho^2}
+
\frac{\PhiN{3}}{\rho^3}.
\end{align}
\end{subequations}
These terms are {\it growing} as $r^3$ for $r \to \infty$, and this is actually essential: their contribution under the reconstruction operator $\S^{\dagger}_{ab}$ combines with $\Lie_{\zeta}g_{ab}$, in \eqref{eq:GHZdecomp}, furthermore leading to a delicate cancelation of various subleading terms and leaving in the end a perturbation with Bondi-like falloff at future null infinity~\cite[Thm. 4]{Hollands:2024iqp}. For this, a specific choice of the growing terms must of course be made; in fact, by working 
through the proof of ~\cite[App. J]{Hollands:2024iqp}, one can see that the entire completion is generated by a single {\it Held Master Scalar} (HMS) $\fHo$ of spin
weight $+2$, defined by
\begin{equation}\label{eq:f seed}
2\,\thpH{}^{3}\,\bar\psi^{5\circ}
=
\edthpH{}^{4}\,\edthH{}^{4}\,\fbHo
-9\,\Psih\Psibh\,\thpH{}^{2}\,\fbHo .
\end{equation}
The four Held integration constants appearing in
\eqref{eq:Phi completion intro} are determined from the HMS $\fbHo$ by the triangular
system
\begin{subequations}\label{eq:Phi n Held}
\begin{align}
\PhiN{3} &= -\thpH\fbHo,
\label{eq:Phi3 Held}\\[2pt]
\PhiN{2} &= -\edthpH{}^2\fbHo,
\label{eq:Phi2 Held}\\[2pt]
\thpH\,\PhiN{1} &= -\tfrac12\,\edthpH{}^{2}\edthH{}^{2}\fbHo
	-\big(2\tabh\,\edthH+\tfrac32\Psih\big)\thpH\fbHo,
\label{eq:Phi1 Held}\\[2pt]
\thpH{}^{2}\,\PhiN{0} &= -\tfrac16\,\edthpH{}^{3}\edthH{}^{3}\fbHo
	-\Big(\tabh\,\edthpH\edthH{}^{2}
	 +\tfrac12\Psih\,\edthpH\edthH\nonumber\\
     &\qquad+\Psih\rhopH\Big)\thpH\fbHo ,
\label{eq:Phi0 Held}
\end{align}
\end{subequations}
noting that 
$\thpH \to -i\omega$ in modes.

Step 3) We build $\zetal,\zetan,\zetam$---which give the residual gauge vector field $\zeta^a$ \cite{Price:2006ke}, by differentiation formulas found, e.g., in \cite[App. E]{Hollands:2024iqp}---from 
\begin{subequations}\label{eq:zeta Held}
\begin{align}
\zetamb &= \edthH\fbHo,
\label{eq:zeta m Held}\\[2pt]
\thpH\,\zetal &= -\Re\Big( \edthH{}^2\fbHo\Big),
\label{eq:zeta l Held}\\[2pt]
\thpH\,\zetan &= -\Re\Big(
	\edthpH\edthH{}^{3}\fbHo
	-\rhopH\,\edthH{}^2\fbHo
	-2\tah\,\edthH\thpH\fbHo\nonumber\\
    &\qquad-\tfrac12\Oh\,\edthH{}^2\thpH\fbHo
	\Big).
\label{eq:zeta n Held}
\end{align}
\end{subequations}
from the HMS $\fbHo$, noting again that 
$\thpH \to -i\omega$ in modes.
In fact, for \eqref{eq:GHZdecomp}, we only need to evaluate not $\zeta^a$ itself but $\Lie_{\zeta}g_{ab}$, which is determined directly by $\zetal,\zetan,\zetam$ through the formulas in App.~\ref{app:Lie}. 

Step 4) We determine $x_{ab}$ from the GHZ transport equations by a backward integration in $r$ from ${\mathscr I}^+$, see, e.g., 
\cite[App. H]{Toomani:2021jlo}, and we determine the zero mode $\dot g_{ab}$ by Abbot-Deser type integrals from $T_{ab}$, see e.g., \cite{Toomani:2021jlo}. These last steps are actually even easier in our model problem as we will see, since the stress tensor is non-zero only locally near the particle. 

We thereby have all pieces in \eqref{eq:GHZdecomp} giving us the metric perturbation in a Bondi-like gauge in IRG.
For completeness, the spin weights of every object appearing in the construction are collected
in Table~\ref{tab:spin-weights-fields}.

\begin{table*}[t]
\caption{Spin weights of the GHP and reconstruction fields entering
the adjusted reconstruction.
The second row is the sign $(-1)^s$ appearing in the conjugation
rule~$\big(\bar X\big)_{\omega,\ell, m}
=
(-1)^{s+\emm}\;\overline{X_{-\omega,\l,-\emm}}$ for mode sum coefficients \eqref{eq:modeexp}.}
\label{tab:spin-weights-fields}
\begin{ruledtabular}
\begin{tabular}{lcccccc}
  & $\fHo$ & $\fbHo$ & $\PhiN{n}$
  & $\zetal,\,\zetan$ & $\zetam$ & $\zetamb$ \\
\hline
  Spin weight $s$
    & $+2$ & $-2$ & $-2$  & $0$ & $+1$ & $-1$ \\
  $(-1)^{s}$
    & $+$  & $+$  & $+$    & $+$ & $-$  & $-$  \\
\end{tabular}
\end{ruledtabular}
\end{table*}

\begin{table*}[t]
\caption{Spin weights of the metric perturbation.}
\label{tab:spin-weights-metric}
\begin{ruledtabular}
\begin{tabular}{lccccccc}
  & $h_{ll}$ & $h_{ln}$ & $h_{nn}$ & $h_{m\bar{m}}$
  & $h_{nm}$ & $h_{n\bar{m}}$ & $h_{mm}$ \\
\hline
  Spin weight $s$
    & $0$ & $0$ & $0$ & $0$ & $+1$ & $-1$ & $+2$ \\
  $(-1)^{s}$
    & $+$ & $+$ & $+$ & $+$ & $-$  & $-$  & $+$  \\
\end{tabular}
\end{ruledtabular}
\end{table*}

\section{Circular orbit in Schwarzschild}
\label{sec:Schw}
Here we outline our scheme for a point particle on a circular orbit in Schwarzschild. Due to the high symmetry, it is convenient to expand any spin-$s$ quantity ${}_s X$ in this spacetime in modes as
\begin{equation}
\label{eq:modeexp}
    {}_s X = \sum_{\omega,\ell,m}  {}_s X_{\omega \ell m}(r) \, {}_s Y_{\ell m}(\theta) \,  e^{-i\omega t+im\varphi}, 
\end{equation}
where our conventions for the spin-weighted spherical harmonics ${}_s Y_{\ell m}(\theta)$  are the same as in \cite{Hollands:2024iqp}. For generic orbits, possible frequencies are the harmonics of the basic orbital frequencies $(\Omega_r, \Omega_\theta, \Omega_\varphi)$ labeled by three integers. For a circular orbit as considered here, this reduces to $\omega=m\Omega,$ with $\Omega  \equiv \Omega_\varphi=\sqrt{M/r_0^3}$, where $r_0$ is the radius of the orbit.

Following \cite{Bourg:2024cgh,Leather:2026gnc}, a puncture $h^{\rm P}_{ab}$ for such an orbit is obtained by the method described in Sec. \ref{sec:comoving punctures}. Then the NP components of the effective stress energy tensor $T^{\rm E}_{ab}$ \eqref{eq:TEdef} are formed using this puncture and the stress energy tensor of the particle \eqref{eq:Tabdef}.  Using a version of the semi-analytic tools developed in \cite{Bourg:2024cgh,Leather:2026gnc}, see Sec. \ref{sec:comoving punctures}, one then obtains the $(\ell,m)$-mode expansion \eqref{eq:modeexp} of the corresponding effective Teukolsky source in \eqref{eq:Teuksource}. Its expansion coefficients are numerically given as an array on a discrete set of $r$-values per mode. The order of the puncture governs the fall-off of the mode amplitudes in $\ell$.

With these at hand, one obtains the $(\ell,m)$ mode expansion coefficients of the perturbed Weyl-scalar $\psi$ \eqref{eq:psidef} by solving the mode version of Teukolsky's equation \eqref{eq:Teuksource}, i.e., the radial Teukolsky equation for the modes coefficients \eqref{eq:modeexp} of $\psi$, with the effective Teukolsky source, and with purely downgoing (at the horizon $r=2M$) and purely outgoing (at future null infinity $r=\infty$) boundary conditions. We have numerically checked that this $\psi$ satisfies peeling \eqref{eq:peeling}, and thereby we can extract the $(\ell,m)$ mode-sum coefficients of the leading coefficient,
\begin{equation}
    \psi^{5\circ} = \sum_{\omega,m,\ell} \psi^{5\circ}_{\omega \ell m} \, {}_{+2} Y_{\ell m}(\theta) e^{-i\omega u + im\varphi},
\end{equation}
by fitting $\psi$ against $r^{-5}$. Recall that  $u=t-r_*$ is the Eddington coordinate, and that quantities with a $\circ$ by construction have no $r$-dependence.
Either by the modal inversion scheme \cite{Ori:2002uv}, or by simply 
integrating $\psi$ four times against $r$ beginning from $r=\infty$ (in practice an $r$-value much larger than both $M$ and the particle's orbital radius $r_0$), we get the minimal solution $\Phi^{\rm min} = O(r^{-1})$ of the inversion relation \eqref{eq:invrel} for the Hertz-potential, in $(\ell,m)$-modes.

Next, we solve the $(\ell,m)$-mode version of our HMS \eqref{eq:f seed}. In doing so, we use that $\thpH = -i\omega$ in modes, and that the combination of angular operators in \eqref{eq:f seed} is precisely the combination appearing in the Teukolsky-Starobinsky identities \cite{TeukolskyPress1974,StarobinskyChurilo1973}---where both observations generalize to Kerr. The $(\ell,m)$-mode coefficients $f^\circ_{\omega \ell m}$ of the HMS are thereby found to be:
\begin{equation}\label{eq:f mode kerr}
\fHo_{\omega\ell m}
=
\frac{2i\,\omega^{3}}{\Dlw}\;\psi^{5 \circ}_{\omega \ell m}
\end{equation}
where $\Dlw$ is a Teukolsky-Starobinsky constant\footnote{
In Kerr, it is given by \eqref{eq:KerrTSconstant}.}. 
Its value in Schwarzschild is 
$\Dlw=\tfrac{1}{16}(\ell-1)^2\ell^2(\ell+1)^2(\ell+2)^2 + 9M^2 \omega^2$.

Using the ladder operator properties of Held's angular 
derivatives in terms of spin-weighted harmonics recalled e.g., in \cite[App. C]{Toomani:2021jlo}, the four Held completion pieces \eqref{eq:Phi n Held} in the Hertz potential are similarly found to be
\begin{subequations}\label{eq:schw closed form}
\begin{align}
\PhiN{0}_{\omega \ell m}
&=
\frac{-\tfrac13 i\omega(-1)^{\emm}}{\Dlw}\\
& \quad \times
\Big[\Bco{2}\Bco{1}-3iM\omega\Bco{1}\Big]
\bar\psi^{5\circ}_{-\omega,\ell,-m},
\label{eq:Phi0 schw}\\[4pt]
\PhiN{1}_{\omega \ell m}
&=
\frac{\omega^{2}(-1)^{\emm}}{\Dlw}
\Big(\Bco{2}-3iM\omega\Big)
\bar\psi^{5\circ}_{-\omega,\ell,-\emm},
\label{eq:Phi1 schw}\\[4pt]
\PhiN{2}_{\omega \ell m}
&=
\frac{2i\omega^{3}\,\Bco{2}/\Bco{1}\,(-1)^{\emm}}{\Dlw}
\bar\psi^{5\circ}_{-\omega,\ell,-m},
\label{eq:Phi2 schw}\\[4pt]
\PhiN{3}_{\omega \ell m}
&=
\frac{-2\omega^{4}(-1)^{\emm}}{\Dlw}
\bar\psi^{5\circ}_{-\omega,\ell,-m},
\label{eq:Phi3 schw}
\end{align}
\end{subequations}
The residual gauge vector $\zeta^a$'s building blocks \eqref{eq:zeta Held} needed to compute $\Lie_\zeta g_{ab}$ by the formulas in App. \ref{app:Lie} are
\begin{subequations}\label{eq:zeta schw}
\begin{align}
\zetamM{\omega \ell m}
&=
\frac{-2i\omega^{3}\sqrt{\Bco{2}/\Bco{1}}}{\Dlw}
\;\psi^{5\circ}_{\omega \ell m},
\label{eq:zeta m schw}\\[4pt]
\zetalM{\omega \ell m}
&=
\frac{\omega^{2}\sqrt{\Bco{2}}}{\Dlw}
\;\psi^{5\circ}_{\omega \ell m}\;+\;{\rm c.c.},
\label{eq:zeta l schw}\\[4pt]
\zetanM{\omega \ell m}
&=
\frac{-\omega^{2}\sqrt{\Bco{2}}}{\Dlw}
\Big(\Bco{1}+\tfrac12\Big)
\;\psi^{5\circ}_{\omega \ell m}\;+\;{\rm c.c.},
\label{eq:zeta n schw}
\end{align}
\end{subequations}
where ``$+\,{\rm c.c.}$'' denotes the mode-space real part,
i.e.\ the addition of $(-1)^{\emm}\bar\psi^{5\circ}_{-\omega,\ell,-\emm}$
inside the same prefactor. We also defined $\sqrt{\Bco{2}}=\tfrac12\sqrt{(\ell-1)\ell(\ell+1)(\ell+2)}$ and 
$\Bco{2}/\Bco{1} = \tfrac12(\ell+2)(\ell-1)$.

In our inward integration scheme, the corrector $x_{ab}$, obtained via the GHZ transport equations, vanishes to the right of the interval $(r_{\rm min}, r_{\rm max})$ around the particle where the puncture is windowed. Thus, outside this window, the reconstructed residual metric \eqref{eq:GHZdecomp} is in this case simply 
\begin{equation}
\label{eq:habadjusted}
    h_{ab}^{\rm R} = 2\Re \S^\dagger_{ab} \Phi - \Lie_\zeta g_{ab} + \dot g_{ab}
    \quad (r>r_{\rm max}).
\end{equation}
The NP components of $\Lie_\zeta g_{ab}$ are obtained by applying the formulas of App. \ref{app:Lie}, which drastically simplify for $a=0$ due to $\tau^\circ = \Omega^\circ = 0$,
mode by mode to \eqref{eq:zeta schw} in automated form using the spin-raising and lowering properties \cite{Chandrasekhar:1983} of Held's angular derivatives as given in Table \ref{tab:Held-coeffs-Kerr}. $\dot g_{ab}$
is the zero mode in our Bondi-like gauge, see e.g., \cite[Eq. C.2]{Hollands:2024iqp}, which is determined in the present case by $\delta M = \mu, \delta a = j/M$, with $\mu$ the particle's mass and $j = \mu \sqrt{M}/\sqrt{r_0-3M}$ the particle's orbital angular momentum.

Since we have the full corrector $x_{ab}$, we of course have the entire metric perturbation \eqref{eq:GHZdecomp} in a mode-decomposed form with coefficients dependent on $r$ {\it all the way} from the horizon to infinity. We will use this to compute the Detweiler redshift at the particle in Sec.  \ref{sec:CircKerr}. When we compute asymptotic quantities, see Sec. \ref{sec:CircKerr} for numerical results (in Kerr), we do not need $x_{ab}$ in the present first order model application of our scheme. On the other hand, at second order, the effective source no longer is supported only near the particle, so the second order corrector is supported all the way from the horizon to infinity and would be needed  for the computation of asymptotic quantities as well.

\section{Circular orbit in Kerr}
\label{sec:CircKerr}

As our main example, we derive the first order metric perturbation in our asymptotically flat IRG for a circular orbit in Kerr. We could proceed by a mode expansion similar to the Schwarzschild case, where spin-weighted spheroidal functions would now appear. While such an ``$(\ell,m)$-mode'' scheme is possible, we here describe an ``$m$-mode'' scheme, which seems more promising in general for inclined, elliptical orbits as needed in reality. However, we will briefly describe 
some changes necessary in order to generalize the $(\ell,m)$-mode scheme to Kerr in Sec. \ref{sec:lmKerr}, as could be useful for moderately inclined and elliptical orbits. 

In the $m$-mode scheme, quantities are generally expanded as 
\begin{equation}
\label{eq:Xmmodes}
    {}_s X = \sum_{\omega,m} {}_s X_{\omega m}(r,\theta) e^{-i\omega t+im\varphi}, 
\end{equation}
as opposed to \eqref{eq:modeexp}. Consequently, the mode amplitudes ${}_s X_{\omega m}(r,\theta)$ in a sense already include an infinite $\ell$-summation. The downside is that, of course, decoupled ordinary differential equations (ODEs) such as the radial and angular Teukolsky equation now become partial differential equations (PDEs) in $(r,\theta)$. The modes therefore have to be stored as an array of number on a 2-dimensional $(r,\theta)$-grid, and the numerical solution is considerably less straightforward. Fortunately, it turns out that the only PDE in our scheme is the Teukolsky equation \eqref{eq:Teuksource}. On the other hand, the GHZ transport equations for the non-zero corrector components $x_{nn}, x_{mn}, x_{m\bar m}$ remain ODEs in $r$ (for each fixed $\theta$-value), and the equation for the HMS $f^\circ$ is an ODE in $\theta$. Furthermore, the inversions needed to find 
the completion pieces from the HMS in Eqs. \eqref{eq:Phi n Held}, \eqref{eq:zeta Held} remain algebraic since $\thpH = -i\omega$ on $m$-modes.

\subsection{Comoving punctures}
\label{sec:comoving punctures}

The puncture is a local, analytically constructed approximation to the singular part of the metric perturbation near the particle's worldline, and it is only ever
needed in a neighborhood thereof cut out by a window step-function in $r$. We use the high-order construction initiated for a circular orbit in Schwarzschild and a scalar toy problem in~\cite{Bourg:2024cgh}, and later generalized to circular orbits in Kerr in the spin-2 case by \cite{Leather:2026gnc}.\footnote{We thank the authors of \cite{Leather:2026gnc} for providing and explaining to us their numerical implementation of their method prior to releasing \cite{Leather:2026gnc}.} It is based on a coordinate system centered on the particle rather
than Boyer--Lindquist coordinates. Let $\gamma$ be the worldline with
proper time $T$ and four-velocity $u^{a}$, and let $e_{\hat a}{}^{a}$ be an
orthonormal tetrad along $\gamma$ whose timelike leg is $e_{\hat 0}{}^{a}=u^{a}$
and whose spatial triad is Fermi--Walker transported. Local spatial
coordinates $(X,Y,Z)$ are obtained by shooting geodesics orthogonal to
$\gamma$ and then Lie-dragging the resulting labels along the congruence,
\begin{equation}\label{eq:comoving conditions}
  u^{a}\nabla_{a}T = 1,
  \qquad
  u^{a}\nabla_{a} (X,Y,Z) = 0,
\end{equation}
so that $(\partial_{T})^{a}=u^{a}$ in a whole neighborhood of $\gamma$ in the coordinates $(T,X,Y,Z)$, and
the particle sits at $(X,Y,Z)=0$ for all $T$.

The payoff of \eqref{eq:comoving conditions} is that comoving stationarity
becomes triviality in coordinates. If a tensor field $F_{a_1\cdots a_p}$ is stationary in the
comoving frame, $\Lie_{u}F_{a_{1}\cdots a_{p}}=0$, then its tetrad components
obey $\partial_{T}F_{\hat a_{1}\cdots \hat a_{p}}=0$, and the field is a function of the
spatial comoving coordinates alone. The local singular field, $h^{\rm P}_{ab}$ of the point
particle, and the particle's stress tensor, $T_{ab}$ \eqref{eq:Tabdef}, are of exactly this type. Its dependence on $(X,Y,Z)$ is fixed by subjecting its local expansion
in $(X,Y,Z)$ to the EEs $\E h^{\rm P}_{ab} = T_{ab}$, 
thereby pushing the information about the particular orbit  into the expansion coefficients. This expansion has the general form in terms of frame index components $\hat a$ for our NP-tetrad $e^a{}_{\hat a} = (n^a, l^a, m^a, \bar m^a)$~\cite{Bourg:2024cgh,Leather:2026gnc}, 
\begin{equation}\label{eq:puncture monomials}
  h^{\rm P}_{\hat a \hat b}
  = \sum_{i,j,k,p,q} c^{\hat a \hat b}_{ijkpq}\,
    X^{i}Y^{j}Z^{k}R^{p}\log^{q}\!R ,
\end{equation}
where $R^{2}=X^{2}+Y^{2}+Z^{2}$ where the constant coefficients $c^{\hat a \hat b}_{ijkpq}$ 
are determined by the EEs. The order of the puncture dictates the number of expansion terms to be computed but it is always finite\footnote{In our case we found that it depends slightly on which metric component one considers, to be sufficient in order to get a sufficiently regular $T_{ab}^{\rm E}$ \eqref{eq:TEdef}. But the result for a scalar in the STF form is $\ell_{\rm max}= 3(n+1)$ where $n$ is the order of the puncture counting from the leading divergence: $1/s$ for first order and $1/s^2$ for second order. Note that the $\log R$ terms only arise at second order.}.
This general structure survives for a generic bound orbit, but
we now specialize to the case of a circular orbit, recording the minimum needed to follow the
$m$-mode reduction; the construction is given in full in
\cite{PZpuncturenotes}, which builds on~\cite{Bourg:2024cgh}, but we give a brief outline in App. \ref{App:genOrbPunct} as forward guidance. 

In order to decompose $h_{\hat a \hat b}^{\rm P}$ into $m$-modes \eqref{eq:Xmmodes}, we require the transformation from body-fixed coordinates $(T,X,Y,Z)$ to Boyer-Lindquist coordinates $(t,r,\theta,\varphi)$ in order to integrate \eqref{eq:puncture monomials} against $e^{-i\omega t+im\varphi}$.
To write these down for our circular orbit, we define
$\varphi_c \defeq \varphi - \Omega t$, which is a corotating phase, and we introduce
\begin{equation}\label{eq:comoving chart}
  X = \frac{r/r_0-1}{\sqrt{\Delta(r_0)}},
  \qquad
  Y = -r_{0}\cos\theta,
  \qquad
  Z = z_c\sin\tfrac{\varphi_c}{2},
\end{equation}
together with a comoving time $T$ that is a fixed linear combination of $t$ and $\varphi$ and employ the  \texttt{Mathematica} \cite{Mathematica} package \emph{xCoba} \cite{xCoba} for coordinate based tensor manipulations. The constants above are fixed by the orbital radius $r_0$, as $\Omega = (a + \sqrt{r_{0}^3/M})^{-1}$, and
\begin{equation}\label{eq:comoving constants 2}
  D_0^2 = \frac{2a\sqrt{M/r_0}+r_0-3M}{r_0},
  \quad
  z_c = \frac{2\sqrt{M \Delta(r_0)}}{\sqrt{r_{0}^3}\,\Omega D_0}.
\end{equation}
Quantities may be pulled back from body-fixed to Boyer--Lindquist coordinates by these formulas.

 Two reductions bring
\eqref{eq:puncture monomials} into a form whose $m$-modes can be evaluated in
closed form. First, powers of $Z$ beyond the first are traded for powers of
$R^{2}-X^{2}-Y^{2}$, which leaves at most one explicit factor of $Z$. Second,
that remaining factor is replaced by the single-valued angular variable
$\xi \defeq z_c\sin\varphi_c$, using
$Z = \tfrac12 \xi + O\!\left(\xi^{3}\right)$ and truncating at the order of
the puncture. The result is
\begin{equation}\label{eq:puncture reduced}
  h^{\rm P}_{\hat a \hat b}
  = \sum_{i,j,k,p,q} d^{\hat a \hat b}_{ijkpq}\,
    X^{i}Y^{j}\xi^{k}R^{p}\log^{q}\!R ,
\end{equation}
now with $k,q\in\{0,1\}$, so that only four types of angular integral can occur.

Indeed, since $\omega = m\Omega$, the $m$-mode amplitudes of \eqref{eq:puncture reduced} can be considered as integrals over $\varphi_c$ at fixed $(X,Y)$. Since $R^{2} = \varrho^{2}_0 + z_c^{2}\sin^{2}(\varphi_c/2)$
with $\varrho^{2}_0 \defeq X^{2}+Y^{2}$ is linear in $\cos\varphi_c$, every such
integral reduces to a single master function of the form
\begin{equation}\label{eq:master J}
  J_{\nu \emm}
  = (R\varrho_0)^{\nu}
    \frac{\Gamma(\nu+1)}{\Gamma(\nu+|\emm|+1)}
    P^{|\emm|}_{\nu}\left( \frac{R^{2}+\varrho^{2}_0}{2R\varrho_0} \right),
\end{equation}
now with $R^{2} = \varrho^{2}_0 +z_c^{2}$, through the dictionary
\begin{subequations}\label{eq:puncture dictionary}
\begin{align}
  R^{p} &\to J_{p/2,\emm},\\
  \xi R^{p} &\to \tfrac{z_c}{2i}\big(J_{p/2,\emm-1}-J_{p/2,\emm+1}\big),\\
  R^{p}\log R &\to \tfrac12\,\partial_{\nu}J_{\nu \emm}\big|_{\nu=p/2},\\
  \xi R^{p}\log R &\to \tfrac{z_c}{2i}\,\partial_{\nu}
   \big(J_{\nu,\emm-1}-J_{\nu,\emm+1}\big)\big|_{\nu=p/2}.
\end{align}
\end{subequations}
The entire $m$-mode puncture is therefore assembled from associated Legendre
functions $P^{|\emm|}_{\nu}$ of a single argument, with no numerical angular quadrature \cite{Bourg:2024cgh}. This is
what makes the puncture side of the $m$-mode scheme cheap enough to be recomputed at every radial grid point.
For generic orbits, part of this structure usefully survives, see App. \ref{App:genOrbPunct}. 

\subsection{Effective source}

Once we have $h^{\rm P}_{ab}$ as in \eqref{eq:puncture reduced}, we multiply it by a radial window function supported in some suitable $(r_{\rm min}, r_{\rm max})$ (which could be varied for numerical verification), and 
{\it afterwards} apply the linearized Einstein-operator, $\E$. This gives us the effective source \eqref{eq:TEdef}, in which the $\delta$-function localized piece corresponding to the particle is cancelled analytically. Since the procedure has already been implemented e.g., by \cite{Bourg:2024vre}, we will not describe it in detail here. After this, we apply Teukolsky's source operator to get the source $j^{\rm T}$ in Teukolsky's equation $\O \psi = j^{\rm T}$, see App. \ref{app:Teuk}.

\subsection{Integration of GHZ equations for corrector}

\label{sec:corrector}

The corrector $x_{ab}$ does not enter the construction of the HMS and completion pieces require for our Bondi-like gauge, but it is required in the full metric perturbation \eqref{eq:GHZdecomp}.  

In IRG, the corrector has three non-vanishing NP components, $x_{nn}, x_{nm}, x_{m\bar m}$. They are found by integrating the GHZ transport equations for them form a triangular
system of radial transport equations along the outgoing principal null
congruence \cite{Green:2019nam}, see e.g., \cite[Sec. 3.1.3]{Hollands:2024iqp}. These can be solved by integration \cite[App. H]{Toomani:2021jlo}. In princple, in our general scheme \cite[Sec. 3.1]{Hollands:2024iqp}, see Sec.~\ref{sec:scheme}, we integrate the GHZ
transport equations inward from ${\mathscr I}^{+}$ with vanishing data there.
In the application considered, the effective source \eqref{eq:TEdef} has compact radial support in the worldtube by construction, so $x_{ab}$ vanishes identically from the outer
edge of the worldtube out to $r=\infty$. In that whole region the metric is
given by \eqref{eq:habadjusted} with no corrector at all\footnote{In our scheme, the corrector is non-zero in the region inside the particle's orbital radius. Since the effective source itself vanishes there, this contribution is pure gauge. It can be removed by a further gauge transformation of the type described in \cite{Toomani:2021jlo}, but this is not considered in the present paper, since our interest is in the asymptotic region.}

The $m$-mode reduction of the GHZ transport equations is again the
substitution \eqref{eq:Xmmodes}, after which each level is an
ordinary integration in $r$ at fixed $z=\cos\theta$. The
two-dimensional problem therefore decouples into $N_{z}$ independent radial
problems, and this --- not the Teukolsky equation --- is the sense in which
the corrector is the easy part of the $m$-mode scheme. Two of the three
levels require a double integration in $r$; the third only a single integration.

Each radial problem is discretized by Chebyshev collocation on two patches
meeting at the particle, with interface conditions imposed on the value and
the derivative, so that the non-smoothness of the windowed effective source at
$r=r_{0}$ is never differentiated across.

\subsection{Weyl scalar and Hertz potential}
\label{sec:weyl and hertz}

This is the one genuinely hard step of the $m$-mode scheme, and the only
place where a PDE has to be solved. 
The Teukolsky wave operator $\mathcal O$ and source $j^{\rm T}$ (obtained from $T_{ab}^{\rm E}$) for $\psi$
take a compact form in the Held-type operators and coefficients of Table \ref{tab:Held-coeffs-Kerr}, which we record in App.~\ref{app:Teuk}, see \eqref{eq:TeukO Held}, \eqref{eq:Teuk T0 Held}. In outgoing Eddington coordinates, in which
$\th = \partial_r$ and $\thpH = \partial_u- (r^2 + a^2 z^2)^{-1} \Delta \partial_r$, the resulting equation is, schematically
\begin{equation}\label{eq:Teuk ABC}
     \hat A \partial_r^2 \psi +2\partial_u \partial_r \psi+ \hat B \partial_r \psi + \hat C \psi = j^{\rm T},
\end{equation}
with coefficients $\hat A$, $\hat B$, $\hat C$ that are obtained by substituting the values of the Held quantities in Table \ref{tab:Held-coeffs-Kerr} into the formulas of
App.~\ref{app:Teuk}. These are \emph{operator-valued} coefficient
functions of $(r,z=\cos \theta)$: only $\hat A = - (r^2 + a^2 z^2)^{-1}\Delta$ is a plain
function of $(r,z)$, while $\hat B$ contains $\Dbar$ and $\Dbar'$ separately, and
$\hat C$ contains them only through the combination $\Dbar\Dbar'$. Note also that
$\hat C$ carries no radial derivatives.

In the $m$-mode reduction we should replace
$\partial_u \to -i\omega$ and $\partial_{\varphi_*}\to i\emm$ in \eqref{eq:Teuk ABC}. The mixed derivative and the $\partial_u$ term inside $\hat C$ then become algebraic,
and
\begin{equation}\label{eq:Teuk ABC mode}
  \hat A\,\partial_r^{2}\psi_{\omega\emm}
  + \big(\hat B_{\omega\emm} - 2i\omega\big)\partial_r \psi_{\omega\emm}
  + \hat C_{\omega\emm}\,\psi_{\omega\emm}
  = j^{\rm T}_{\omega\emm}.
\end{equation}
Because $\Dbar$ and $\Dbar'$ still carry $\partial_z$ by Table \ref{tab:Held-coeffs-Kerr}, \eqref{eq:Teuk ABC mode}
remains a PDE in $(r,z)$ at each
$(\emm,\omega)$. 

We solve it with retarded\footnote{By this, we mean purely downgoing boundary conditions at the horizon and purely outgoing ones at $r=\infty$. This corresponds to the retarded solution in physical space.} boundary
conditions. In the future, we intend to use a multidomain spectral method~\cite{PanossoMacedo:2024pox,Cownden:2026msq},
with domain boundaries placed at the edges of the worldtube so that the
windowed effective source is smooth on each domain. However, in the present analysis, we use standard $(\ell,m)$-mode Teukolsky solvers in this step \cite{BHPTeukolsky, Lo:2023fvv,Yin:2025kls}. We then verify a
posteriori that the solution satisfies peeling \eqref{eq:peeling}, so that
$\psi^{5\circ}$ may be read off as the coefficient of $r^{-5}$.

By the inversion
relation \eqref{eq:invrel}, the Hertz potential $\Phi$ is obtained by integrating $\bar\psi$ inward in $r$ four
times --- in modes, $\bar\psi$ at $(\emm,\omega)$ is assembled from $\psi$ at
$(-\emm,-\omega)$ by the conjugation rule quoted with
Table~\ref{tab:spin-weights-fields} ---
starting from a radius much larger than both $M,r_0$ and
taking vanishing data there. This selects precisely the minimal solution
$\Phi^{\rm min} = O(r^{-1})$, and the growing in $r$ pieces that are left undetermined by
\eqref{eq:invrel} are exactly the Held integration constants
$\PhiN{0},\dots,\PhiN{3}$ supplied by the completion scheme of Sec.~\ref{sec:scheme}, where we should concretely substitute the entries of Table \ref{tab:Held-coeffs-Kerr} into \eqref{eq:Phi n Held} with $\thpH \to -i\omega$, $\partial_{\varphi_*} \to im$.

\subsection{HMS and residual gauge vector}
\label{sec:mmode HMS}

For this completion, and for the residual gage vector $\zeta^a$ in \eqref{eq:GHZdecomp}, we next require the HMS $f^\circ$, determined by 
\eqref{eq:f seed}. In $m$-modes, this is an 8th order ODE in $z=\cos \theta$. If we do not want to perform an $(\ell,m)$-mode reduction, which would render this equation algebraic by \eqref{eq:f mode kerr}, we need to solve this ODE. Intrinsically, the HMS $f^\circ$ is smooth, i.e., it is smooth in the sense of a mathematical object called a section in a ``GHP/Held-bundle'', see e.g., \cite[Sec. 2.1]{Hollands:2024iqp} for explanation. With our conventions of coordinates and NP tetrads, this corresponds to certain boundary conditions at the poles. We solve our ODE by
collocation on Legendre--Gauss--Lobatto nodes in $z$, subject to these boundary conditions. Two ingredients make this practical.

The first is pole factorization. For a spin-weight-$s$ GHP/Held quantity $X$ at azimuthal
number $\emm$ intrinsic smoothness translates into a certain (singular) endpoint behavior in our NP-frame. It is convenient to factor this behavior out and work throughout with
the regular reduced amplitude $\hat X$ defined by
\begin{equation}\label{eq:pole factor}
  X \;=\; (1-z)^{|\emm+s|/2}\,(1+z)^{|\emm-s|/2}\;\hat X .
\end{equation}
In this representation Held's angular derivatives (see Table \ref{tab:Held-coeffs-Kerr}) act as first-order
operators with polynomial coefficients, of the general form
\begin{equation}\label{eq:reduced eth}
  \edthH \;\to\; \tfrac{1}{\sqrt2}\Big[\,C_{+}(\emm,s,z)\, D_{z}
   + B_{+}(\emm,s,z) - a\omega\,C_{+}(\emm,s,z)\Big],
\end{equation}
and similarly for $\edthpH$ with coefficients $C_{-},B_{-}$ and the opposite
sign on the undifferentiated term. Here $C_{\pm}$ and $B_{\pm}$ are
polynomials of degree at most one in $z$, fixed by the signs of $\emm\pm s$,
and $D_z$ is the Legendre differentiation matrix
representing $\partial_z$. Because
the removed factor in \eqref{eq:pole factor} it vanishes
at the poles, no quantity in the chain ever has to be evaluated at a
coordinate singularity\footnote{The spin weight entering \eqref{eq:reduced eth}, which must correspond to the one included in the definition of the operator according to Table \ref{tab:Held-coeffs-Kerr}, is always the
correct one.}.

The second is that \eqref{eq:f seed} is solved as a dense linear system
rather than integrated. The left-hand side is assembled by eight successive
applications of \eqref{eq:reduced eth}, four of $\edthH$ and four of
$\edthpH$; applying this composite operator to each collocation basis vector
in turn gives the $N_{z}\times N_{z}$ matrix column by column (here $N_z$ is the number of collocation points). The system is
row- and column-equilibrated, factorized by full-pivot LU (a method that decomposes a matrix into a product of a lower triangular matrix and an upper triangular matrix to enhance numerical stability), and refined
iteratively; the relative residual and the condition number of the
equilibrated matrix are returned alongside the solution and are the primary
diagnostics of the step.

With $\hat{f}^{\circ}_{\omega\emm}$ in hand, the four
Held integration constants \eqref{eq:Phi n Held} and the components of
$\zeta^{a}$ \eqref{eq:zeta Held} are formed on the same grid by further
applications of \eqref{eq:reduced eth}. Since $\Phi$ with the growing completion pieces is a cubic polynomial in $\rho^{-1}$ \eqref{eq:rho def} and
$\S^{\dagger}_{\hat a \hat b}$ is second order in $r$-derivatives, each NP component of the
adjusted residual field perturbation is returned as 
\begin{equation}\label{eq:laurent}
  h^{\rm R}_{\hat a \hat b} \;=\; A^{(1)}_{\hat a \hat b}\,r + A^{(0)}_{\hat a \hat b}
  + O\left( \frac{1}{r} \right),
\end{equation}
with the two Laurent polynomial coefficients stored as arrays over the $z$ nodes, separately
for the reconstructed piece, the gauge piece, and their sum. The Bondi
fall-off is then simply the statement that $A^{(1)}_{\hat a \hat b}=A^{(0)}_{\hat a \hat b}=0$ at every
collocation node for the $nn$, $nm$ and $mm$ NP-tetrad components\footnote{All other NP-components vanish identically by construction.}, which can be checked
pointwise without any mode sum.

\subsection{Numerical results for $h_{ab}^{\rm R}$}
\label{sec:mmode results}

We begin this section with benchmarks of the $m$-mode chain. So that every
intermediate object can be compared against a ``closed-form'' reference, the runs
reported here were first performed at $a=0$: the operators, the grid and the
reconstruction are the Kerr ones described above, but the background is
Schwarzschild, where the $(\ell,m)$ sums are available in closed form, see Sec. \ref{sec:Schw}. All figures use a circular geodesic at $r_{0}=10M$ with $M=1$, fitted $\psi$ data summed through $\ell=20$, and $N_{z}=21$ Legendre--Gauss--Lobatto (LGL) nodes in $z=\cos\theta$.

\begin{figure*}
  \includegraphics[width=\textwidth]{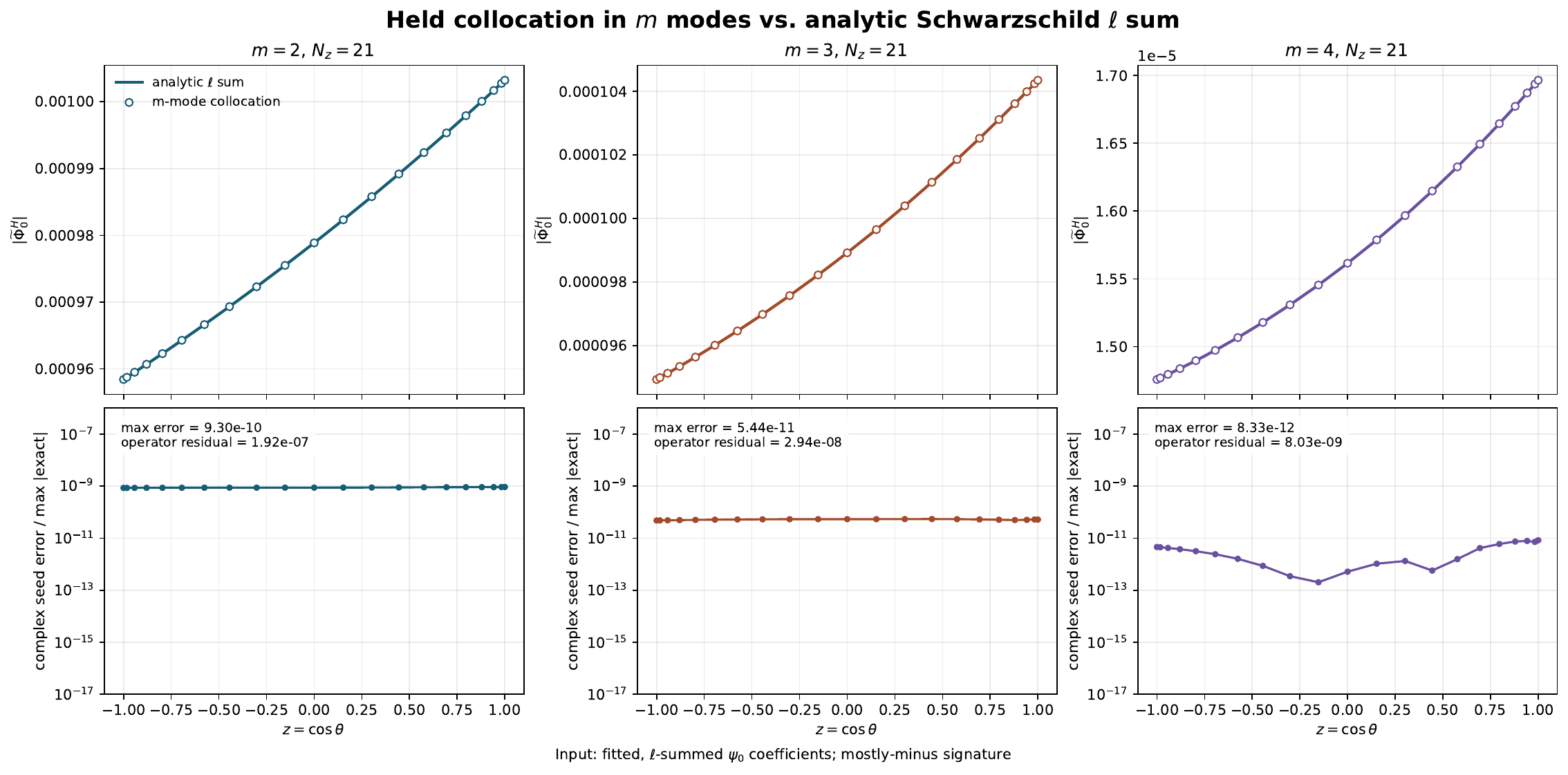}
  \caption{Validation of the $m$-mode HMS for $a=0$. Top row: the modulus of
 ${\Phi}^{0\circ}$ on the $N_{z}=21$ LGL grid
  (circles), obtained by solving \eqref{eq:reduced eth} in $z=\cos \theta$, against the
  analytic Schwarzschild $\ell$ sum (solid). Bottom row: the pointwise complex
  error relative to $\max|\text{exact}|$. The quoted operator residual is the
  relative residual of the reduced angular equation when the analytic form is
  substituted, and measures the discretization rather than the solve.}
  \label{fig:mmode seed}
\end{figure*}

\emph{Finding the HMS $f^\circ$.} Figure~\ref{fig:mmode seed} compares the collocation solution of the $m$-mode scheme of Sec. \ref{sec:mmode HMS} for the HMS with the $(\ell,m)$-scheme in Sec. \ref{sec:Schw} ($\ell$-summed HMS) for $m=2,3,4$. The maximum relative complex HMS
errors are $9.30\times10^{-10}$, $5.44\times10^{-11}$ and
$8.33\times10^{-12}$, with relative operator-equation residuals for the
analytic HMS of $1.92\times10^{-7}$, $2.94\times10^{-8}$ and
$8.03\times10^{-9}$. The errors fall steeply with $m$, as expected from the
pole factorization \eqref{eq:pole factor}: the residual function $\hat X$
becomes smoother as $|m|$ grows, so the spectral representation converges
faster. The equilibrated condition number of the discrete operator falls
correspondingly, from $8.6\times10^{7}$ at $m=2$ to $4.9\times10^{5}$ at
$m=4$, which is why the equilibrated full-pivot solve with iterative
refinement is used rather than a plain factorization. This validates both the
$\ell$-summed input and the $m$-mode seed inversion.

\begin{figure*}
  \includegraphics[width=\textwidth]{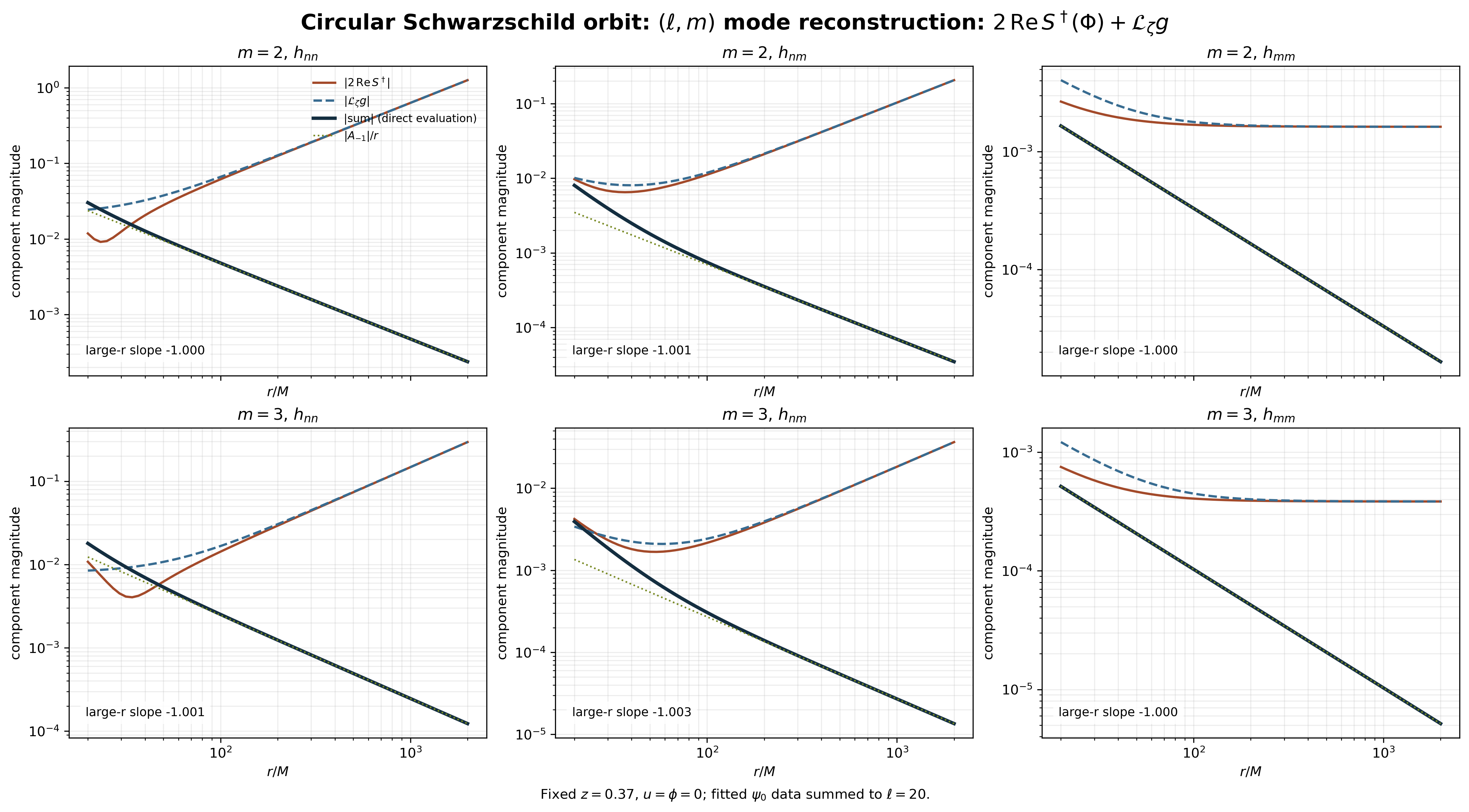}
  \caption{$(\ell,m)$ mode scheme reference for the adjustment
  \eqref{eq:habadjusted}, built from the closed-form Schwarzschild $(\ell,m)$
  components summed at fixed $m$ through $\ell=20$, evaluated at
  $z=\cos \theta=0.37$, $u=\varphi=0$. Each panel shows the reconstructed piece
  $|2\Re\S^{\dagger}_{\hat a \hat b}\Phi|$, the gauge piece $|\Lie_{\zeta}g_{\hat a \hat b}|$, their sum, and the fitted slope coefficient.  We clearly see that the individually
  growing pieces cancel to leave a $1/r$ tail.}
  \label{fig:modal falloff}
\end{figure*}

\emph{Cancellation and Bondi falloff.} The central claim of
Sec.~\ref{sec:scheme} is that the linear in $r$ and constant asymptotic expansion orders carried separately
by $2\Re\S^{\dagger}_{\hat a \hat b}\Phi$ and by $\Lie_{\zeta}g_{\hat a \hat b}$ cancel in the
sum \eqref{eq:habadjusted}, leaving a perturbation with Bondi-like falloff (here hatted indices $\hat a$ indicate NP-frame components).
Figure~\ref{fig:modal falloff} exhibits this cancellation in the $(\ell,m)$-mode scheme. Across $1254$ nonstationary radiative mode components through
$\ell=20$, the largest relative cancellation error in the growing and constant
Laurent coefficients is $4.90\times10^{-16}$, i.e.\ at the level of double
precision. Direct finite-radius sums at $z=0.37$ for $m=2,3$ show $h_{nn}$,
$h_{nm}$ and $h_{mm}$ approaching $1/r$ between $r=20M$ and $r=2000M$, with
fitted large-radius slopes between $-1.003$ and $-1.000$. The residual drift
from $-1$ is the expected contamination from the subleading $\log r/r^{2}$ and
$1/r^{2}$ terms at finite radius, and shrinks as the fitting window is pushed
outward.

\begin{figure*}
  \includegraphics[width=\textwidth]{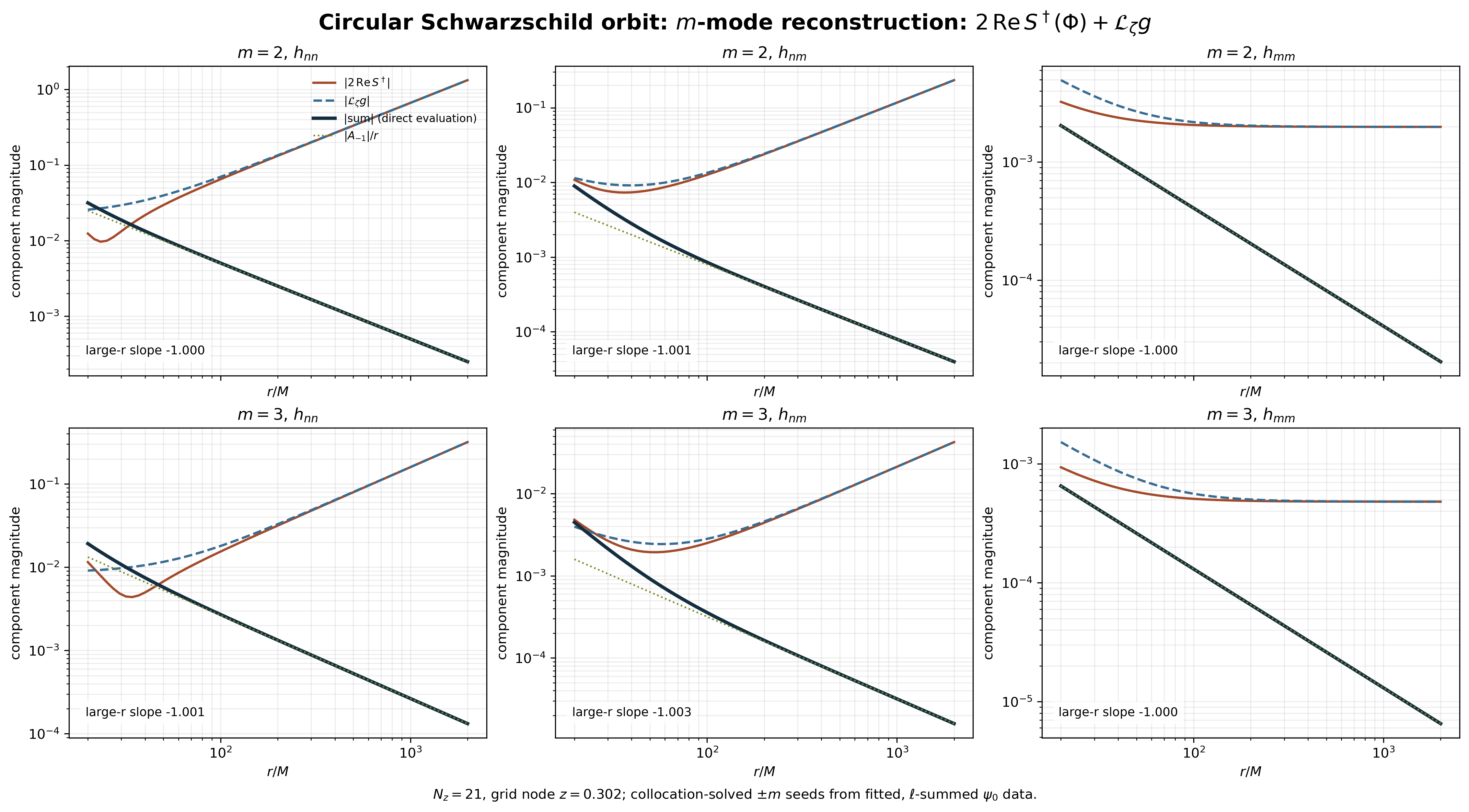}
  \caption{The same six components as Fig.~\ref{fig:modal falloff},
  constructed instead from the $m$-mode scheme on the $N_{z}=21$ grid and evaluated at the interior node $z=\cos \theta = 0.3020$. The
  reconstruction forms $\Phi^{0\circ}$ through $\Phi^{3\circ}$, applies the
  spectral Held angular derivatives for $\S^{\dagger}_{\hat a \hat b} \Phi$ and $\Lie_{\zeta}g_{\hat a \hat b}$,
  and sums the complex Laurent coefficients directly at each radius, without
  ever projecting back onto individual $(\ell,m)$ modes.}
  \label{fig:spectral falloff}
\end{figure*}

\emph{$m$-mode scheme.} Figure~\ref{fig:spectral falloff} repeats
the radial test with the full $m$-mode pipeline in place of the modal
reference: the $+m$ and $-m$ adjustment scalars are obtained by collocation, the reflected
partner being the conjugate of the reduced $-m$ adjustment scalars since the angular
harmonics already supply the odd-$m$ reflection phase, and all subsequent
angular derivatives are taken spectrally. At the interior node
$z=0.3020$ the six plotted components again have large-radius slopes between
$-1.003$ and $-1.000$. Across all interior nodes the largest relative
difference from the modal reference of Fig.~\ref{fig:modal falloff} is
$1.04\times10^{-6}$ for $m=2$ and $1.54\times10^{-7}$ for $m=3$, while the
largest relative residual in the growing and constant Laurent coefficients is
$2.24\times10^{-12}$. The cancellation is therefore not an artifact of the
mode sum: it survives the spectral realization of the Held operators, which is
the only form available in Kerr.

\begin{figure*}
  \includegraphics[width=\textwidth]{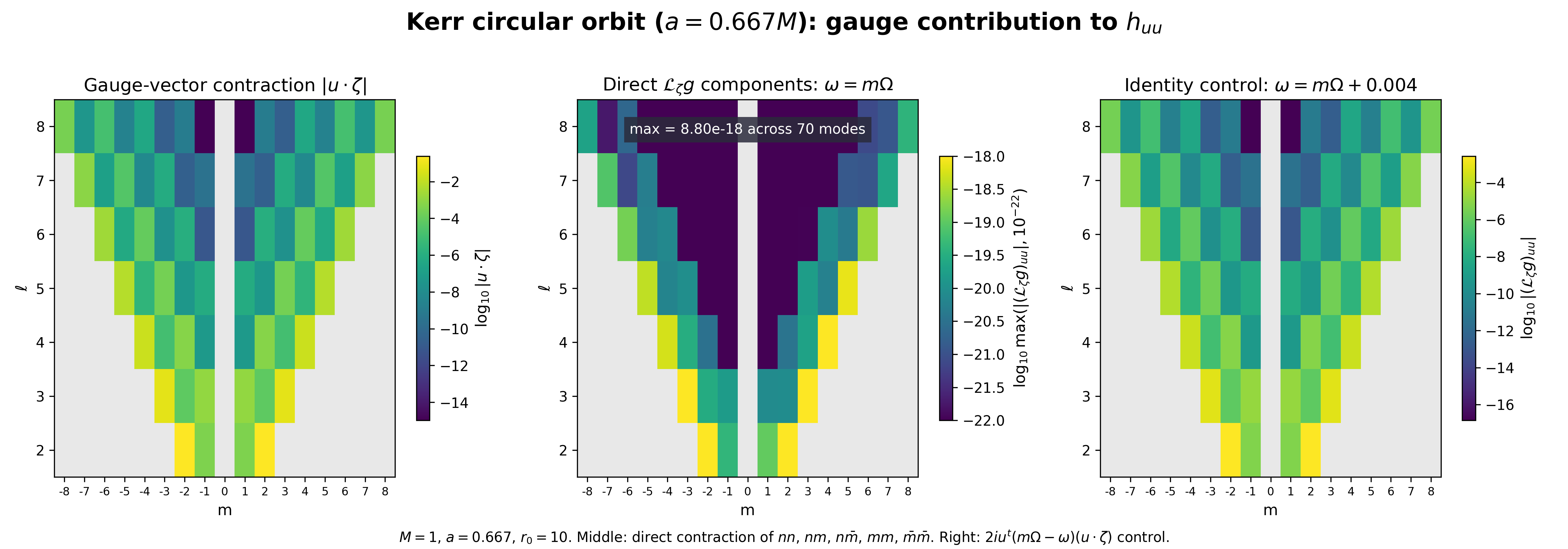}
  \caption{Gauge vector $\zeta^a$ contribution to $h_{ab}^{\rm R} u^a u^b$ in \eqref{eq:GHZdecomp} for the Kerr ($a=0.667M$) circular
  orbit ($r_0=10M$), over $70$ modes with $2\le\ell\le8$ and $m\neq0$. Left: the worldline
  contraction $|u^a\zeta_a|$, which is generically nonzero. Center: the
  helical pullback $|\Lie_{\zeta}g_{ab} u^{a}u^{b}|$ evaluated on shell at
  $\omega=m\Omega$.  Right is the control quantity $2i(u^b t_b)(\omega-m\Omega)(u_a\zeta^a)$, with $\omega$ detuned. Color scales are $\log_{10}$ and differ
  between panels.}
  \label{fig:lie zeta,Kerr}
\end{figure*}

\emph{Detweiler redshift.} 
This is among the simplest observables in the self-force context at linear order. Denoting the (non-normalized) circular orbit's tangent by $u^a = (\partial_t)^a + \Omega(\partial_\varphi)^a$, the first order gauge invariant redshift observable relative to a distant observer tangent to $t^a=(\partial_t)^a$ is \cite{Detweiler:2008ft,Barack:2018yvs} $\delta z=\frac{1}{2} h^{\rm R}_{ab} u^a u^b$ in a gauge such that $\Lie_u h_{ab}^{\rm R}=0$. It is invariant under gauge transformations $\Lie_X g_{ab}$ such that $[X,u]^a=0$ \cite[App. B]{Detweiler:2008ft}. In particular, this holds for 
each mode of our gauge vector field $\zeta^a$. In fact, using only the modal time and angular dependence, one can easily see that each mode contributes 
\begin{equation}
    u^a u^b \Lie_\zeta g_{ab} \to 
    -2i t^a u_a (m\Omega-\omega) \zeta^b u_b     
\end{equation}
which clearly vanishes as $\omega=m\Omega$. However, this is not manifestly obvious when we determine $\Lie_\zeta g_{ab} u^a u^b$ through the formulas in App. \ref{app:Lie}. But 
Fig.~\ref{fig:lie zeta,Kerr} confirms this: over $70$ modes the maximum
helical value is $2.42\times10^{-19}$, at roundoff, while $|u^a \zeta_a|$
ranges over some twelve decades in the same modes. Detuning the frequency by
$\delta\omega=0.004$ raises the maximum to $3.93\times10^{-3}$, confirming
that the null result stems from $m\Omega=\omega$, and not an accidental cancellation or a quantity that was never computed.

The remaining non-gauge pieces alone in $h_{ab}^{\rm R}$ in 
our decomposition \eqref{eq:GHZdecomp} therefore give the Detweiler redshift. 
The values we obtained for $h^{\rm R}_{ab}u^a u^b$ are:
$-0.2160630(3)$ ($a=0$)
and $-0.19847(1)$ ($a=0.667M$),
in agreement with the no-string GHZ field \cite{Bourg:2024vre} 
and a traditional radiation-gauge (CCK) metric-reconstruction calculation
\cite{vanDeMeentShah2015}.
In Fig.~\ref{fig:partial sum,redshift} we show the approach of the partial $\ell$-sums in the calculation of  $h^{\rm R}_{ab}u^a u^b$ to the reference value: namely Eq.~\eqref{eq:GHZdecomp} times $u^a u^b$ 
when decomposing the terms ``$(2\Re \S^\dagger_{ab} \Phi + x_{ab})u^a u^b$" into $(\ell,m)$-modes and summing first over $\ell:|m|\to \ell_{\rm max}$, then over $m:-20\to 20$ and then adding a numerically-fitted tail for $|m|>20$.

\begin{figure*}
  \includegraphics[width=0.45\textwidth]{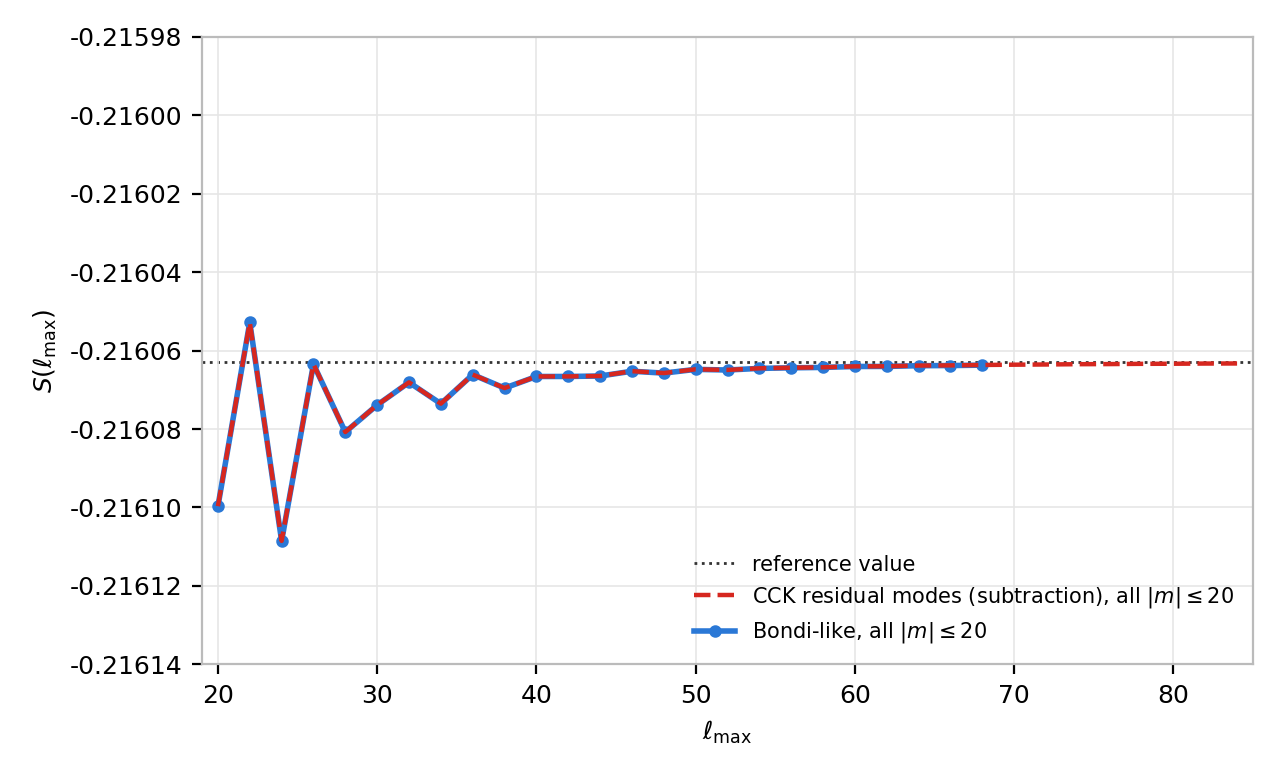}
  \includegraphics[width=0.45\textwidth]{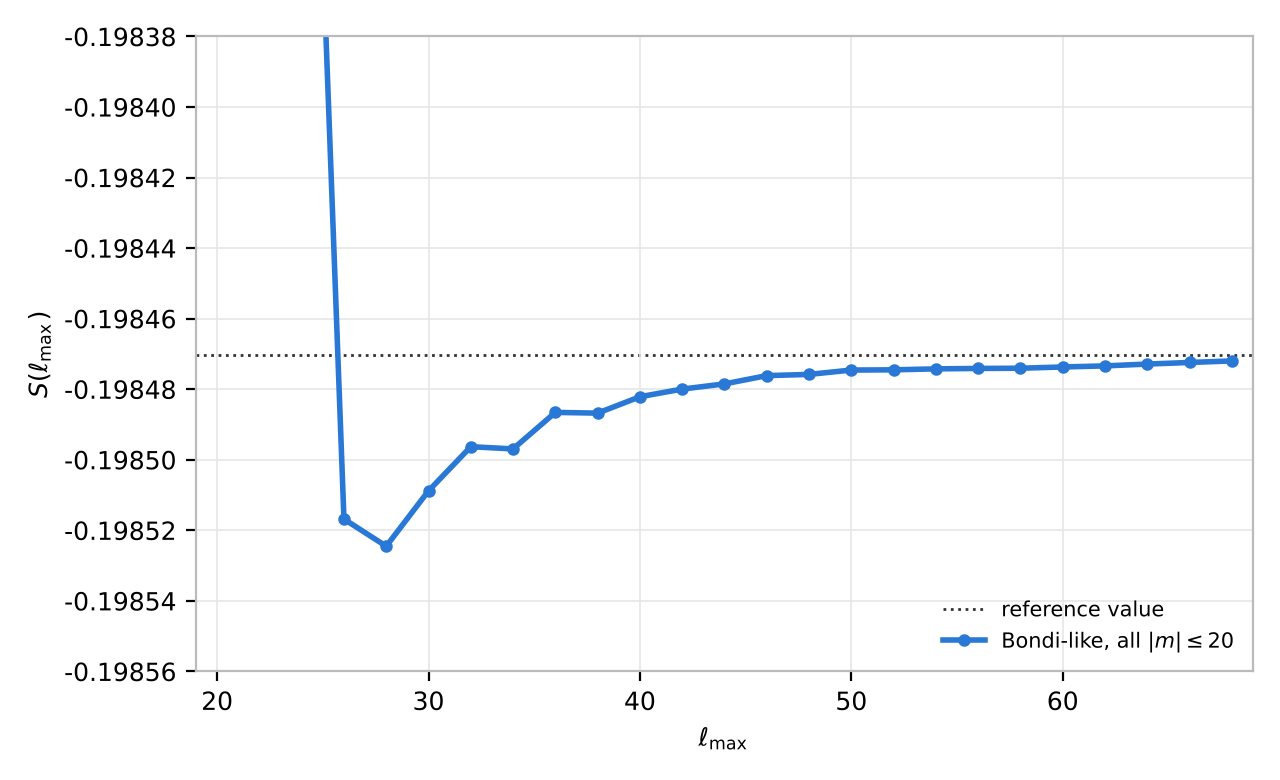}
  \caption{
Partial $\ell$-sums  in the calculation of  $h^{\rm R}_{ab}u^a u^b$ for $a=0$ (left) and $a=0.667M$ (right), $r_0=10M$: $S(\ell_{\rm max})$ is  Eq.~\eqref{eq:GHZdecomp} 
when decomposing the terms ``$(2\Re \S^\dagger_{ab} \Phi + x_{ab})u^au^b$" into $(\ell,m)$-modes and summing first over $\ell:|m|\to \ell_{\rm max}$, then over $m:-20\to 20$ and then adding a numerically-fitted tail for $|m|>20$.
  The horizontal dashed lines are the reference values
  $h^{\rm R}_{ab}u^a u^b=-0.216063$ $(a=0)$ and $-0.198470$ $(a=0.667M)$.
  The dashed orange curve in the left plot corresponds to the calculation using the CCK residual modes (up to $\ell_{\max}=68$ it lies under the Bondi-like curve: absolute difference $\le4.7\times10^{-8}$).}
  \label{fig:partial sum,redshift}
\end{figure*}

\begin{figure*}
  \includegraphics[width=\textwidth]{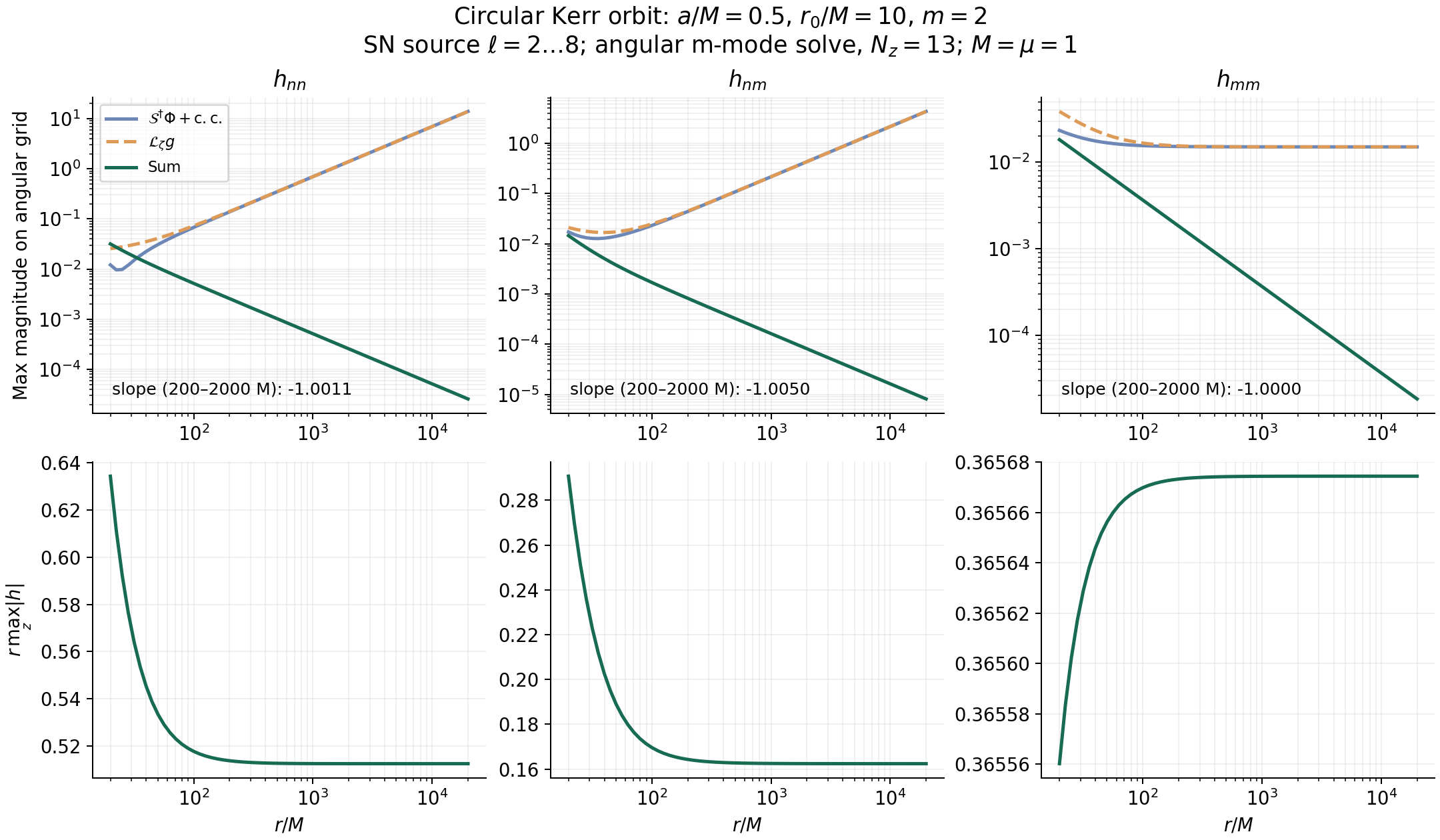}
  \caption{The NP-components of the 
  adjusted metric perturbation $h^{\rm R}_{\hat a \hat b}$ \eqref{eq:GHZdecomp}
  in the asymptotic region $r \to \infty$,
  constructed from the $m$-mode scheme on the $N_{z}=21$ grid and evaluated at the interior node $z=\cos \theta=0.3020$ for a moderately spinning Kerr black hole $(u=\varphi=0)$. The top row again shows the cancellation between the reconstructed piece and the gauge piece, leaving only the subleading $1/r$ decay.  The second row shows the extrapolation of the $1/r$-pieces, which can be seen as observables for gravitational radiation. Shown are absolute values.}
  \label{fig:spectral falloff Kerr}
\end{figure*}

\emph{Asymptotic quantities in Kerr.} We have also validated the cancellation at non-decay orders $r^j, j=0,1$ between the reconstructed piece $2 \Re \S^\dagger_{ab} \Phi$ and the gauge piece $-\Lie_\zeta g_{ab}$ in the full metric perturbation \eqref{eq:GHZdecomp} as $r \to \infty$ in Kerr. This is shown in Fig. \ref{fig:spectral falloff Kerr}.  

\begin{figure*}
  \includegraphics[width=\textwidth]{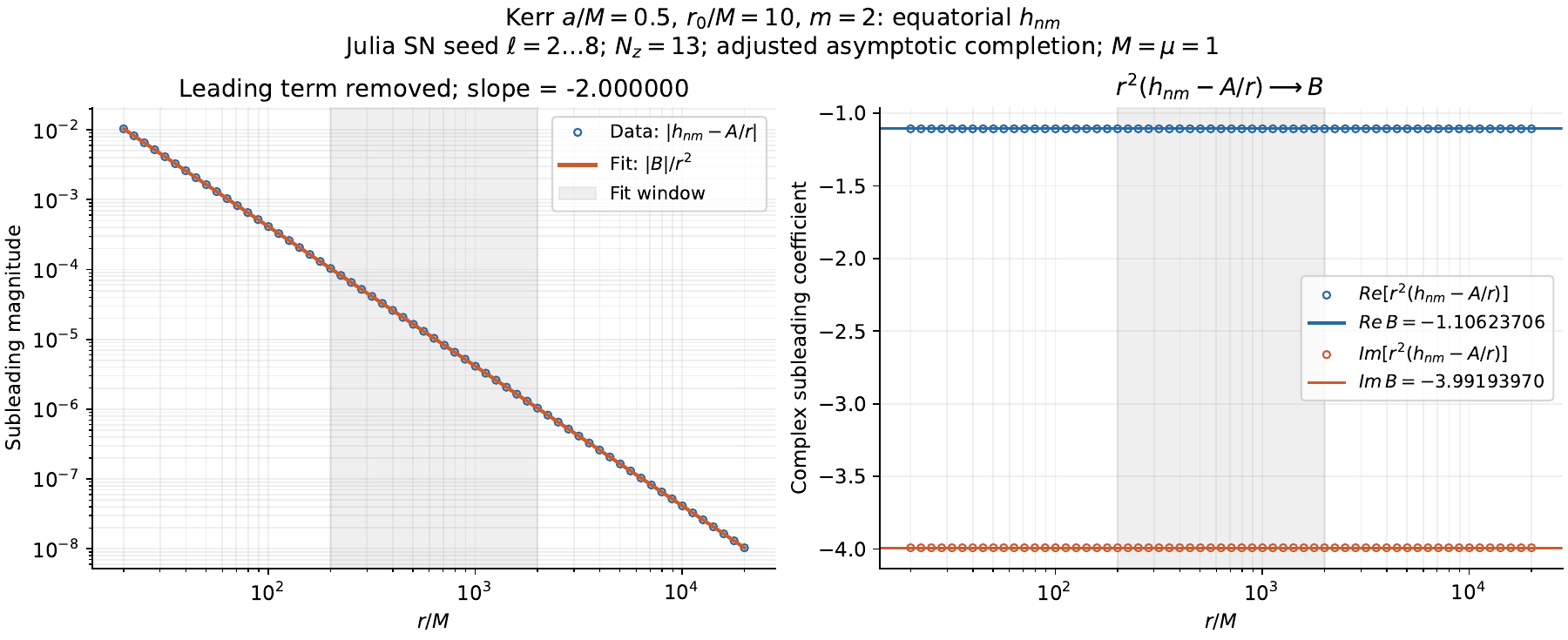}
  \caption{Subleading $1/r^2$ piece of $h_{nm}$, as required for the Bondi angular momentum aspect.}
  \label{fig:spectral falloff Kerr1}
\end{figure*}

For example, as shown in \cite[App. D]{Hollands:2024iqp}, the Bondi mass aspect is given up to normalization by $\lim_{r \to \infty} rh_{nn}$
in outgoing Eddington coordinates, as could be inferred for a particluar angle $\theta$ and for the $m=2$ mode contribution in the first panel 
in the second row of Fig. \ref{fig:spectral falloff Kerr}.  Likewise, $\omega \lim_{r \to \infty} r\Re(h_{mm})$ gives the ``$+$'' entries of the Bondi-news tensor, whereas 
$\omega \lim_{r \to \infty} r\Im(h_{mm})$ gives the ``$\times$'' entries for each particular mode with dependence $e^{-i\omega t+im\varphi}, \omega = m\Omega$. These quantities hence directly follow from the third panel 
in the second row of Fig. \ref{fig:spectral falloff Kerr}.

The Bondi angular momentum aspect can be retrieved from $h_{nm}$ at the order $1/r^2$ by formulas given in \cite[App. D]{Hollands:2024iqp}. Since the leading order is $1/r$, as follows from the second panel 
in the second row of Fig. \ref{fig:spectral falloff Kerr}, another step is necessary to dig out the this subleading piece. It is shown in Fig. \ref{fig:spectral falloff Kerr1}. That plot suggests also the absence of any terms of the form $(\log r)/r^2$ in the asymptotic expansion.\footnote{The absence of such terms would follow automatically for a smooth $\sI^+$ in the perturbed spacetime. It cannot be taken for granted for our source which exists infinitely far into the past.} 

\subsection{Remarks on $(\ell,m)$-mode scheme in Kerr} 
\label{sec:lmKerr}

For moderately inclined and eccentric orbits an $(\ell,m)$-mode treatment
remains attractive, since every step of our scheme other than the Teukolsky
solver, and the GHZ transport equations for the corrector \cite{Green:2019nam} stays algebraic. Furthermore, the GHZ transport equations can be solved by $r$-integration per $(\ell,m)$-mode \cite[App. H]{Toomani:2021jlo}, which is no substantial difference to solving them per angle $\theta$ in the $m$-mode scheme, and thus would be expected to be of a comparable computational cost. We summarize here what changes in the remaining steps of our scheme relative to the
Schwarzschild case of Sec.~\ref{sec:Schw}.

A generic bound geodesic has three fundamental frequencies $\boldsymbol{\Omega}=(\Omega_\varphi, \Omega_r, \Omega_\theta)$ as we recall in App. \ref{app:gen what is lost}. Spin-weighted quantities ${}_s X$, such as the Weyl scalar $\psi$, are correspondingly expanded in terms of their harmonics, $\omega = \boldsymbol{k} \cdot \boldsymbol{\Omega}$, $\boldsymbol{k} = (k_\varphi, k_r, k_\theta) \in {\mathbb Z}^3$ and in 
spin-weighted spheroidal eigenfunctions as
\begin{equation}\label{eq:modeexp generic}
  {}_s X = \sum_{\ell \boldsymbol{k}} {}_s X_{\omega \ell m}(r)\,
  {}_s S_{\omega \ell m}(\theta) \,
  e^{-i\omega u + im\varphi_*},
\end{equation}
in place of \eqref{eq:modeexp}, 
where $m=k_\varphi$, and $\omega = \boldsymbol{k} \cdot \boldsymbol{\Omega}$. For $\psi$, the $(\ell,m)$ mode amplitudes ${}_{+2}\psi_{\omega \ell m}$ are in fact the quantities returned by most production Teukolsky frequency domain codes and the asymptotic amplitude
$\psi^{5\circ}_{\omega \ell m}$ is the leading $1/r^5$ piece of this quantity (which can be viewed as an asymptotic quantity at $\scri$ if we want).

The associated modal HMS $f^\circ_{\omega \ell m}$ is produced algebraically as in Schwarzschild by \eqref{eq:f mode kerr} in the spheroidal basis, where $D_{\omega \ell m}$ is 
now the Teukolsky-Starobinsky constant for Kerr, see e.g., \cite[Lem. 2.7]{CasalsTeixeiraDaCosta2021}:
\begin{equation}
\label{eq:KerrTSconstant}
\begin{split}
      \Dlw =& 
  \tfrac{1}{16} \Bigl[ \lambdabar^2 (\lambdabar + 2)^2 + 40 a\omega \lambdabar^2 (m - a\omega) \\
  &+ 48 a\omega \lambdabar (m + a\omega) + 144 (a\omega)^2 (m - a\omega)^2 \Bigr]\\ 
  &+ 9 M^2 \omega^2,
\end{split}
\end{equation}
where $\lambdabar=\lambdabar_{\ell m}(a\omega)$ is the 
the angular eigenvalue of the spheroidal harmonics in the conventions of 
\cite[Eqs. 2.3, 2.1]{CasalsTeixeiraDaCosta2021}.
On the other hand, Held's angular derivatives, which are required in our scheme, act algebraically on spin-weighted \emph{spherical} harmonics, see e.g., \cite[App. C]{Toomani:2021jlo}.

If we want to avoid giving up on the expansion in spheroidal harmonics and keep the Held-operators algebraic, then we have to use a well-known expansion
\begin{subequations}\label{eq:spheroidal to spherical}
\begin{align}
  {}_s S_{\omega \ell m}(\theta)
  &= \sum_{L} {}_s b^{\omega \ell m}_{L}\, {}_s Y_{Lm} (\theta),
  \label{eq:b coeffs}\\
 \Longrightarrow \fHo_{\omega L m}
  &= \sum_{\ell} \fHo_{\omega \ell m}\, {}_{+2} b^{\omega \ell m}_{L} ,
  \label{eq:f Y basis}
\end{align}
\end{subequations}
see, e.g., \cite{Berti:2005gp,kavanagh2016analytical}.
The mixing coefficients $b^{\omega \ell m}_{L}$ are computed once per mode and cached,
since every Held chain downstream reuses them. We denote  the half-width at which the expansion is truncated as $N_{\rm mix}$,
\begin{equation}\label{eq:mixing truncation}
{}_s b^{\omega \ell m}_{L}\ \text{retained for}\ |L-\ell|\le N_{\rm mix}.
\end{equation}

In Kerr, the Held operators $\edthH$ and $\edthpH$ (Table \ref{tab:Held-coeffs-Kerr}) are not simply spin-raising/lowering operators but acquire off-diagonal pieces proportional to
$a\omega$, see e.g., \cite[App. C]{Toomani:2021jlo}. Furtheromore, the multiplication operators $\tah$, $\tabh$ and $\Oh$---which vanish identically when $a=0$---are themselves off-diagonal. However, all five, when considered as matrix operators,
are tridiagonal in $L$ at fixed $\emm$.
Since these operators are linear $\cos\theta$ and $\sin\theta$, they involve in fact only 
``nearest neighbor'' couplings, such as 
\begin{equation}\label{eq:cos theta rec}
\cos\theta\; {}_s Y_{L,m}
=
c_{L+1}\, {}_s Y_{L+1,m}
+b_{L}\, {}_s Y_{L,m}
+c_{L}\, {}_s Y_{L-1,m},
\end{equation}
with
\begin{equation}\label{eq:cos theta coeffs}
{}_s c_{Lm}=\sqrt{\frac{(L^{2}-\emm^{2})(L^{2}-s^{2})}{L^{2}(4L^{2}-1)}},
\quad
{}_s b_{Lm} =-\frac{s\,\emm}{L(L+1)},
\end{equation}
and similarly for $\sin\theta \, {}_s Y_{Lm}$. 
Hence $\cos \theta$ corresponds to a banded tri-diagonal matrix $\mathsf C$ and $\sin \theta$ to a banded tri-diagonal matrix $\mathsf S$, respectively. Then 
$\Held \Omega = -2 i a \mathsf C$,
$\Held \tau = -\frac{ia}{\sqrt2} \mathsf S$
etc. The $L = 0$ case is degenerate and must be considered separately. 
%We give these expression in appendix \ref{}? \PZ{Should we???}

The multiplication operator $\rho^{-1}$ \eqref{eq:rho def} needed for the completion pieces is likewise represented by the tri-diagonal matrix
\begin{equation}\label{eq:R matrix}
\mathsf R_{L\ell} = -r\,\mathbf{1} + i a\,\mathsf C_{L\ell},
\qquad
\mathsf C_{L\ell} = \text{matrix of \eqref{eq:cos theta rec}},
\end{equation}

The mode amplitudes of the completion piece of the Hertz potential \eqref{eq:Phi completion intro} are, when considered as a vector in $\ell$ at fixed $\boldsymbol{k}$,
\begin{equation}\label{eq:Phi adj vector}
\boldsymbol{\Phi}^{\rm adj}
=
\boldsymbol{\Phi}^{0\circ}
+\mathsf R\,\boldsymbol{\Phi}^{1\circ}
+\mathsf R^{2}\,\boldsymbol{\Phi}^{2\circ}
+\mathsf R^{3}\,\boldsymbol{\Phi}^{3\circ}.
\end{equation}
Since each of these five angular operators is tridiagonal, a chain of $p$ of them is banded with half-bandwidth $p$. The longest chain is the
$\edthpH^{\,3}\edthH^{\,3}$ appearing in $\PhiN{0}$, and
\eqref{eq:Phi0 Held} adds at most three further neighbours, so
$\Phiadj$ at output multipole $L$ requires the HMS out to $\ell \le L+6$,
and the HMS in turn requires spheroidal amplitudes out to
$\ell\le L+6+N_{\rm mix}$, where $N_{\rm mix}$ is the half-width at which
\eqref{eq:b coeffs} is truncated. The cost of the completion
is then $O(L_{\max})$ per $\boldsymbol{k}$, which is still negligible when compared with the radial Teukolsky solve that produce its input.

The upshot is that nothing structural in the improved reconstruction scheme is lost with an $(\ell,m)$ scheme in going to Kerr and to generic orbits. The map from the Teukolsky amplitudes $\psi^{5\circ}_{\omega \ell m}$ to
the completed Hertz amplitudes $\boldsymbol{\Phi}^{\rm adj}$ remains purely
algebraic, mode by mode, with no further radial or angular ODEs to solve beyond
those already required for $\psi$ itself. What is diagonal in $(\ell,m)$ for $a=0$ now is merely \emph{banded} via two independent mechanisms, the spheroidal-to-spherical transformation \eqref{eq:b coeffs}, which couples a window of half-width $N_{\rm mix}$, and the five Held operators which are tridiagonal in $\ell$ at fixed $\emm$. Both truncations are numerically under control: the mixing coefficients decay rapidly in $|L-\ell|$ at fixed $a \omega$, so that a modest $N_{\rm mix}$ suffices and the fixed $L+6$ buffer inherited from the longest Held chain
$\edthpH^{\,3}\edthH^{\,3}$ is exact rather than approximate. As an implementation, one simply solves the radial Teukolsky equation out to
$\ell \le L_{\max}+6+N_{\rm mix}$ and monitors the convergence of
$\boldsymbol{\Phi}^{\rm adj}$ under increasing $N_{\rm mix}$. 

\medskip
\noindent
{\bf Acknowledgements:} We thank Patrick Bourg for discussions and explanations of the implementation of \cite{Leather:2026gnc}. We thank Adam Pound, Andrew Spiers, and Barry Wardell for discussions.

\medskip
\noindent
{\bf AI assistance declaration:}
During the preparation of this work, the authors used Claude's Opus (Anthropic) to assist with numerical investigations of the Detweiler redshift quantity in Sec. \ref{sec:mmode results}. Specifically, Claude was prompted to apply existing code referred to in App. \ref{sec:ref map} to produce numerical results and to generate the corresponding Fig. \ref{fig:partial sum,redshift}. In addition, we used Claude's Opus for generating documentation, as well as plotting and unit testing scripts. All scientific interpretation of the results, as well as the writing of the manuscript, were carried out solely by the authors. The authors take full responsibility for the content of this publication.

\appendix

\section{Lie derivative of $g_{ab}$ along $\zeta^{a}$}
\label{app:Lie}
Since $\zeta^a$ is by construction compatible with the traceless IRG, we have $\Lie_{\zeta}g_{ab}l^{a}=0$ and $\Lie_{\zeta}g_{ab}m^{a}\bar m^{b}=0$. 
The remaining, non-zero, NP components are \cite{Price:2006ke}
\begin{widetext}
\begin{equation}
\begin{split}
\left(\Lie_{\zeta}g \right)_{mm} 
&=
-2\bar\rho\,\edthH{}^2\zetal
+2\edthpH\edthH\fHo
+2\Oh\thpH\fHo
+4\bar\rho\,\tah\,\edthpH\fHo
+8\rhopH\fHo ,
\label{eq:Lie mm}
\end{split}
\end{equation}

\begin{equation}
\begin{split}
\left(\Lie_{\zeta}g\right)_{nm}
&=
\bar\rho\,\edthpH\edthH{}^2\zetal
-2\rho\bar\rho\,\tabh\,\edthH{}^2\zetal
+\rho(\rho+\bar\rho)\Psih\,\edthH\zetal
-\bar\rho\,\Oh\,\edthpH{}^2\edthH\fHo
+2\rho\,\tabh\,\edthpH\edthH\fHo
\non\\
&\quad
+\Big(\frac{1}{\bar\rho}
-2\bar\rho\,\Oh{}^{2}\Big)\,\edthpH\thpH\fHo
+2(\rho+\bar\rho)\Oh\tabh\,\thpH\fHo
-2\bar\rho\,\tah\,\edthpH{}^2\fHo
-\big(\rho^{2}\Oh+\rho\big)\Psih\,\edthpH\fHo
\non\\
&\quad
+\bar\rho\,\Psibh\,\edthpH\fHo
-6\bar\rho\,\Oh\rhopH\,\edthpH\fHo
+4\rho\bar\rho\,\tah\tabh\,\edthpH\fHo
+8\rho\rhopH\tabh\,\fHo ,
\label{eq:Lie nm}
\end{split}
\end{equation}

and

\begin{equation}
\begin{split}
\left(\Lie_{\zeta}g \right)_{nn}
&=
\Re\Bigg[
4\rho\bar\rho\,\tabh\,\edthpH\edthH{}^2\zetal
-4\rho^{2}\bar\rho\,\tabh{}^{2}\,\edthH{}^2\zetal
+2\rho^{2}\Psih\,\edthpH\edthH\zetal
+4\rho^{2}\bar\rho\,\Psih\tabh\,\edthH\zetal
\non\\
&\qquad
-2\edthpH{}^3\edthH\fHo
-2\Big(\frac{1}{\rho}+4\Oh\Big)\,\edthpH{}^2\thpH\fHo
-4\rho\bar\rho\,\Oh\tabh\,\edthpH{}^2\edthH\fHo
+4\big(4+\bar\rho\,\Oh-\rho\bar\rho\,\Oh{}^{2}\big)\tabh\,\edthpH\thpH\fHo
\non\\
&\qquad
-\Big(8\rhopH+3\rho\Psih+3\bar\rho\Psibh+2\rho^{2}\Oh{}^{2}\Psih
+8\rho\bar\rho\,\tah\tabh\Big)\edthpH{}^2\fHo
+4\rho^{2}\tabh{}^{2}\,\edthpH\edthH\fHo
+4\rho(\rho+2\bar\rho)\Oh\tabh{}^{2}\,\thpH\fHo
\non\\
&\qquad
+8\rho^{2}\bar\rho\,\tah\tabh{}^{2}\,\edthpH\fHo
+4\rho\bar\rho\,\Psih\tabh{}^{2}\,\edthpH\fHo
-24\rho\bar\rho\,\Oh\rhopH\tabh\,\edthpH\fHo
+16\rho^{2}\rhopH\tabh{}^{2}\,\fHo
\Bigg].
\label{eq:Lie nn}
\end{split}
\end{equation}

The remaining components follow by conjugation,
$\Lie_{\zeta}g_{ab}n^{a}\bar m^{b}=\overline{\Lie_{\zeta}g_{ab}n^{a}m^{b}}$
and
$\Lie_{\zeta}g_{ab}\bar m^{a}\bar m^{b}=\overline{\Lie_{\zeta}g_{ab}m^{a}m^{b}}$,
which in mode space are 
\begin{equation}\label{eq:Lie conj modes}
\big(\Lie_{\zeta}g\big)_{n\bar m, \omega \ell m}
= -(-1)^{\emm}\,\overline{
\big(\Lie_{\zeta}g\big)_{nm, -\omega \ell -m}
},
\qquad
\big(\Lie_{\zeta}g\big)_{\bar m\bar m,\omega \ell \emm}
= (-1)^{\emm}\,\overline{\big(\Lie_{\zeta}g\big)_{mm,-\omega \ell -m}} .
\end{equation}
\end{widetext}
The relative minus sign between the two is the $(-1)^{s}$ arising when conjugating a spin-weighted harmonic.

\section{Teukolsky operators in Held form}
\label{app:Teuk}

The Teukolsky wave operator $\O$ for the $0$-perturbed Weyl scalar, denoted $\psi$ in this paper, appears in\eqref{eq:Teuksource}, 
which we rewrite as $\O \psi = j^{\rm T}$, where $j^{\rm T}=\S^{ab}T_{ab}$ is Teukolsky's source \eqref{eq:Teuksource}. Especially when working with $(\ell,m)$-modes, it is convenient to express these quantities in terms of \eqref{eq:rho def} and Held's coefficients and operators from Table \ref{tab:Held-coeffs-Kerr}, because these are related to spin-raising and -lowering operators, see e.g., \cite[App. C]{Toomani:2021jlo}.
\begin{widetext}
\begin{equation}\label{eq:TeukO Held}
\begin{split}
\half \mathcal O \psi
&= \half\,\rho\bar\rho\,(3\PsiH\rho + 8\rhopH)\,\psi
 - \half\Big[\PsiH\rho\,(3\rho+\bar\rho) + 2\bar\rho\,\rhopH
   + \rho\bar\rho\Big(-\PsibH + 6\rho\,\tauH\taubH
   + (\rhopH+\rhopbh)\,\Oh\Big)\Big]\,(\th\psi)
\non\\
&\quad
+ \th\thpH\psi
+ \rho\bar\rho\,\taubH\,(\th\Dbar\,\psi)
+ \rho\bar\rho\,\tauH\,(\th\Dbar'\psi)
- (4\rho+\bar\rho)\,(\thpH\psi)
- \rho\bar\rho\,(\Dbar\Dbar'\psi),
\end{split}
\end{equation}
and the source is
\begin{equation}\label{eq:Teuk T0 Held}
\begin{split}
j^{\rm T} \equiv \S^{ab} T_{ab} &= - \bar{\rho}^2 \Dbar^2 T_{ll} + 2 \bar{\rho} \Dbar \th T_{lm} - \th^2 T_{mm} + 2 \bar{\rho}^2 \left( 2 \rho + \bar{\rho} \right) \tah \Dbar T_{ll} - 2 \bar{\rho} \left( 2 \rho - \bar{\rho} \right) \tah \th T_{lm}\\
& - 2 \bar{\rho} \left( 2 \rho + \bar{\rho} \right) \Dbar T_{lm} + 2 \left( 2 \rho + \bar{\rho} \right) \th T_{mm} - 2 \bar{\rho}^3 \left( 2 \rho + \bar{\rho} \right) \tah{}^2 T_{ll} + 4 \bar{\rho}^3 \tah T_{lm} - 4 \rho \bar{\rho} T_{mm}.
\end{split}
\end{equation}
\end{widetext}
Important structural relations in the Held formalism are $[\th,\Dbar]=0=\th(\tah,\Oh,\rhopH,\Psih)$ \cite{Held:1974,Held:1975}, which could be used e.g., to rewrite \eqref{eq:Teuk T0 Held} in equivalent forms.

\section{Remarks on punctures for generic orbits}
\label{App:genOrbPunct}

\subsection{What is lost, and what is not}
\label{app:gen what is lost}

In this appendix, we outline how one could generalize the methods by \cite{Leather:2026gnc} used for the punctures in this work, to more general orbits.
First, for a generic orbit, the comoving chart \eqref{eq:comoving conditions} conditions are untouched. 

Second, the singular field---and therefore neither the residual field---is no longer exactly stationary in the comoving
coordinates; i.e,
\begin{equation}\label{eq:Lu nonzero}
  \Lie_{u}h^{\rm P}_{ab}\neq0
  \quad\Longleftrightarrow\quad
  \partial_{T}h^{\rm P}_{\hat a \hat b}\neq0 ,
\end{equation}
so that the NP tetrad components, labeled by hatted indices $\hat a$, retain a dependence on comoving time,
where $T$ agrees with proper time on the worldline; $T\vert_\gamma = \tau$.

Third, the map between the comoving coordinates $(T,X,Y,Z)$ and the background Boyer-Lindquist coordinates $(t,r,\theta,\varphi)$, in terms of which the modes of the  reconstruction algorithm are defined, ceases to be linear with constant coefficients and becomes quasi-periodic.

\subsection{Orbital phases for generic bound orbits}
\label{app:gen phases}

Generic geodesics satisfy the radial and polar equations,
\begin{equation}\label{eq:geodesic rtheta}
  \Sigma^{2}\Big(\frac{\dd r}{\dd\tau}\Big)^{2} = \mathcal R(r),
  \qquad
  \Sigma^{2}\Big(\frac{\dd\theta}{\dd\tau}\Big)^{2} = \Theta(\theta),
\end{equation}
 coupled only through the
common factor $\Sigma^{2}=(r^2+a^2 \cos^2 \theta)^2$. Mino time $\lambda = \int \dd\tau/\Sigma$ decouples them into independent periodic
sectors with torus frequencies
$\Upsilon_{\theta}, \Upsilon_{r}$. See e.g., \cite{Pound:2021qin}, for a detailed review of these and other concepts of orbital motion in Kerr and for further references. The associated phases
\begin{equation}\label{eq:mino phases}
  q_{r} = \Upsilon_{r}\lambda,
  \qquad
  q_{\theta} = \Upsilon_{\theta}\lambda,
\end{equation}
 advance uniformly in Mino time, not in proper time.

The $t$- and $\varphi$-geodesic equations have right-hand sides that are sums of a
function of $r$ alone and a function of $\theta$ alone, and integrate to
a secular piece plus an oscillatory piece,
\begin{equation}\label{eq:secular split}
  t_0(\lambda) = \Upsilon_{t}\lambda + \Delta t(\boldsymbol{q}),
  \qquad
  \varphi_{0}(\lambda) = \Upsilon_{\varphi}\lambda + \Delta\varphi(\boldsymbol{q}),
\end{equation}
where $\qb\defeq(q_{r},q_{\theta})$ and $\Delta t$, $\Delta\varphi$ have zero
average over the torus. The frequencies conjugate to Boyer--Lindquist time,
which are the ones entering the mode decomposition, are the ratios of the
secular rates,
\begin{align}\label{eq:Omega from Upsilon}
  \Omega_{\alpha} = \frac{\Upsilon_{\alpha}}{\Upsilon_{t}},
  \quad \alpha\in\{r,\theta,\varphi\},
\end{align}
where
\begin{align}
  \omega \equiv \omega_{\boldsymbol{k}} = 
  \boldsymbol{\Omega} \cdot \boldsymbol{k},
\end{align}
with $\boldsymbol{k} = (k_\varphi, k_r, k_\theta) \in {\mathbb Z}^3$ are the harmonics of the orbital frequencies appearing in the mode-decompositions.

\subsection{The local expansion on the torus}
\label{app:gen local}

The comoving coordinates \eqref{eq:comoving chart} are built from $r_{0}$,
$\theta_{0}$ (the particle's initial radius and latitude) now
depend on the phases,
\begin{equation}\label{eq:comoving generic}
  X = X(\boldsymbol{x};\qb),
  \quad
  Y = Y(\boldsymbol{x};\qb),
  \quad
  Z = Z(\boldsymbol{x};\qb),
\end{equation}
still with the particle at $X=Y=Z=0$. Here $\boldsymbol{x}=(\varphi_c, \theta, r)$, 
with the corotating phase now defined as
\begin{equation}\label{eq:psi generic}
  \varphi_c = \varphi - \varphi_{0}(\lambda).
\end{equation}
 Unlike for circular orbits, $\varphi_{c}$ is generally no longer linear in $t$ due to the oscillatory part of \eqref{eq:secular split} that this introduces.

The monomial structure \eqref{eq:puncture monomials} is
unchanged. What changes is that the coefficients, together with \eqref{eq:comoving constants 2},
become functions on the orbital two-torus,
\begin{equation}\label{eq:coeffs on torus}
  c^{\hat a\hat b}_{ijkpq}\to c^{\hat a\hat b}_{ijkpq}(\qb),
\end{equation}
and $z_{c}\to z_{c}(\qb)$ etc. Each is then expanded as a Fourier
series on the torus,
\begin{equation}\label{eq:torus series}
  c^{\hat a\hat b}_{ijkpq}(\qb)
  = \sum_{\kb\in\mathbb Z^{2}}
    c^{\hat a\hat b}_{ijkpq,\kb}\,e^{-i\kb\cdot\qb},
  \quad
  \kb\cdot\qb \defeq nq_{r}+kq_{\theta}.
\end{equation}

\subsection{The master integral}
\label{app:gen master}

For a circular equatorial orbit $R^{2}$ is an even function of $\varphi_c$, which
reduces every $\varphi_c$ integral to the single master function
\eqref{eq:master J}. The evenness is due to the discrete isometry $(t,\varphi)\to(-t,-\varphi)$, and composing it with a time
translation gives an isometry that maps the
circular orbit to itself.

For a generic orbit this reflection maps the orbit to its time reverse; the invariance is lost and an odd term appears. The most general dependence a single azimuthal angle can
generate is then
\begin{equation}\label{eq:Rsq generic}
  R^{2} = a(\qb) - b(\qb)\cos\varphi_c - c(\qb)\sin\varphi_c .
\end{equation}
Since $b\cos\varphi_c+c\sin\varphi_c = \beta\cos(\varphi_c-\delta)$ with
$\beta\defeq(b^{2}+c^{2})^{1/2}$ and $\tan\delta\defeq c/b$, this is again a
single cosine, merely displaced: the point of closest approach on the $\varphi_c$
circle sits at $\varphi_c=\delta$ rather than at $\varphi_c=0$. The substitution
$\chi=\varphi_c-\delta$ in the defining integral then costs only a phase,
\begin{equation}\label{eq:J generic}
  J_{\nu\emm}(\qb) = e^{-i\emm\delta(\qb)}\,
  J_{\nu\emm}\big(a(\qb),\beta(\qb)\big),
\end{equation}
so that \eqref{eq:master J} continues to do all the work.

\subsection{From $m$-modes to frequency modes}
\label{app:gen two layers}

The frequency modes are assembled in two steps. First, at fixed orbital
phase, the local azimuthal transform
%\begin{equation}\label{eq:local m mode}
%  \widetilde h^{\,{\rm P}}_{ab;m}(\qb;r,\theta)
%  = \frac{1}{2\pi}\int_{-\pi}^{\pi}
%    h^{\rm P}_{ab}(\qb;\boldsymbol{x})\,e^{-im\varphi_c}\,\dd\varphi_c ,
%\end{equation}
which is the only place \eqref{eq:J generic} and the dictionary
\eqref{eq:puncture dictionary} enter. %Because $\varphi_c$ and $\varphi$ differ by
%a shift, the index $\emm$ here is already the azimuthal harmonic index of the
%final mode, and no mixing in $\emm$ occurs.

Second, the transform over the torus. Using \eqref{eq:secular split} and
\eqref{eq:Omega from Upsilon},
\begin{equation}\label{eq:phase split}
  e^{-im\varphi_{0}}e^{\,i\omega t}
  = e^{\,i\kb\cdot\qb}\,
    e^{\,i\omega \Delta t(\qb)}\,
    e^{-i\emm\Delta\varphi(\qb)} ,
\end{equation}
with $m=k_\varphi, \omega = \boldsymbol{\Omega} \cdot \boldsymbol{k}$
since $\omega \Upsilon_{t}- m\Upsilon_{\varphi}
= k_r \Upsilon_{r}+k_\theta \Upsilon_{\theta}$. Averaging over $t$ and changing
variables to the Mino phases gives
\begin{widetext}
\begin{equation}\label{eq:torus transform}
  h^{\rm P}_{ab;\boldsymbol{k}}(r,\theta)
  = \frac{1}{(2\pi)^{2}}\int_{0}^{2\pi}\!\!\!\int_{0}^{2\pi}
    \widetilde h^{\,{\rm P}}_{ab;m}(\qb;r,\theta)\,
    \mathcal J(\qb)\,
    e^{\,i\omega\Delta t(\qb)\,-\,im\Delta\varphi(\qb)}\,
    e^{\,i(k_r q_{r}+k_\theta q_{\theta})}\,
    \dd q_{r}\,\dd q_{\theta},
  \qquad
  \mathcal J \defeq \frac{1}{\Upsilon_{t}}\frac{\dd t}{\dd\lambda} .
\end{equation}
\end{widetext}
Each factor has a definite role. The phase $e^{-im\Delta\varphi}$ is related to \eqref{eq:secular split}, reflecting the fact that the azimuth against $\varphi_{0}$ 
itself oscillates; $e^{\,i\omega\Delta t}$ corrects the mismatch between the
orbital clock and Boyer--Lindquist time; $e^{\,i(k_rq_{r}+k_\theta q_{\theta})}$
performs the harmonic extraction; and $\mathcal J$ is the Jacobian of the
change from the $t$-average that defines the Fourier coefficient to the
uniform average over the Mino phases. %Since
%$\dd t/\dd\lambda = \Upsilon_{t}+\partial_{\lambda}\Delta t$ is biperiodic
%rather than constant, $\mathcal J$ averages to unity over the torus but does
%not drop out mode by mode.

In the circular limit $\Delta t=\Delta\varphi=0$, $\mathcal J=1$, and
$\widetilde h^{{\rm P}}_{ab;m}$ is independent of $\qb$; the integral
collapses to $\delta_{k_r,0}\delta_{k_\theta,0}$ and the single helical frequency
$\omega=m\Omega_{\varphi}$ is recovered. The generic-orbit algorithm is
therefore the circular one of Sec.~\ref{sec:comoving punctures}, evaluated on
a grid in $(q_{r},q_{\theta})$ and followed by \eqref{eq:torus transform}.

\section{Reference map to numerical code}\label{sec:ref map}

For the reader interested in using our open source
\href{https://github.com/PZ808/GHZ_numeric}{code},
we have written a comprehensive \href{https://github.com/PZ808/GHZ_numeric/blob/master/Notes/Bondi_GHZ_paper_code_map.md}{README} file which indicates which header files, classes, and functions in the code are associated with generating the main quantities in each section, as well as a general description for its use.

\bibliographystyle{apsrev4-2}
\bibliography{MyReferences}
\end{document}